\documentclass[a4paper,12pt]{article}
\usepackage{amsmath,enumerate}
\usepackage{graphicx}
\usepackage{multirow}
\usepackage{xcolor}
\usepackage{minitoc}
\usepackage{braket}
\usepackage{subfig}
\usepackage{slashed}
\usepackage{float}
\usepackage{xcolor}
\usepackage{cite}
\usepackage{datetime}
\definecolor{MyBlue}{rgb}{0.25,0.5,0.75}
\usepackage[colorlinks=true,allcolors=blue]{hyperref}
\newcommand{\lapprox}{%
	\mathrel{%
		\setbox0=\hbox{$<$}
		\raise0.6ex\copy0\kern-\wd0
		\lower0.65ex\hbox{$\sim$}
}}
\newcommand{\gapprox}{%
	\mathrel{%
		\setbox0=\hbox{$>$}
		\raise0.6ex\copy0\kern-\wd0
		\lower0.65ex\hbox{$\sim$}
}}
\newdateformat{monthyeardate}{%
	\monthname[\THEMONTH], \THEYEAR}
\begin{document}

\begin{center}

{\Large \bf Study on the global minimum and $H\to\gamma\gamma$
in the Dirac scotogenic model}\\[20mm]

Raghavendra Srikanth Hundi\\
Department of Physics, Indian Institute of Technology Hyderabad,\\
Kandi - 502 284, India.\\[5mm]
E-mail: rshundi@phy.iith.ac.in \\[20mm]

\end{center}

\begin{abstract}

We have analyzed the vacuum structure of the Dirac scotogenic model, whose
scalar sector consists of two complex Higgs doublets and a real singlet
field. In this model, the standard model like Higgs doublet acquires non-zero
vacuum expectation value (VEV), whereas, the other two fields acquire zero
VEVs. This pattern of VEVs constitute a minimum, which is the desired vacuum
of the model. After analyzing the scalar potential of this model, we have
found that other vacua are also possible in this model. We have shown that
plenty of parameter space exist where the desired vacuum of this model
is the global minimum. We have studied the implications of scalar sector of
this model on the observable quantity of signal strength of Higgs to diphoton
decay. After evaluating this quantity, we have found that the current
experimental values of this quantity can be fitted in this model. Lastly,
we have studied on the possibility of making any of the additional scalar fields
of this model as a candidate for dark matter.

\end{abstract}
\newpage

\section{Introduction}\label{s1}

With the discovery of Higgs boson at the LHC \cite{ATLAS:2012yve,CMS:2012qbp},
the spontaneous breaking of electroweak
symmetry has been verified. In the standard model (SM), the breaking of this symmetry is
explained by postulating a single scalar Higgs doublet \cite{Higgs:1964ia,Higgs:1964pj,
Higgs:1966ev,Englert:1964et,Guralnik:1964eu}, which acquires non-zero
VEV at the minimum of the scalar potential. A consequence of this
breaking mechanism is the existence of the Higgs boson, which is found in the LHC.
As of now, the properties of the Higgs boson, which are measured at the LHC, agrees
with the SM predictions \cite{Workman:2022ynf}. However, more
work is to be done in order to precisely measure the couplings of Higgs boson to all
the SM particles. On the other hand, several reasons exist for the extension
of SM \cite{Quigg:2004is,Ellis:2009pz}. As a result of this, it is worth to explore theories
by proposing additional
Higgs doublets. A minimal extension to the SM, in this aspect, is the two
Higgs doublet model (2HDM) \cite{Lee:1973iz}, where the field content is as same that of SM
apart from an extra scalar Higgs doublet.

In the SM, at the minimum of the scalar potential, the VEV of the
scalar doublet breaks only the electroweak
symmetry. In contrast to this, in the 2HDM, both the Higgs doublets can acquire
VEVs in such a way that, in addition to the electroweak symmetry, $CP$ and charge
symmetries can also be broken spontaneously \cite{Ferreira:2004yd,Barroso:2005sm}.
Different forms of VEVs to the Higgs
doublets are possible in the 2HDM, due to the parameter choice of the model.
As a result of this, in the 2HDM, the possible vacua are categorized as follows
\cite{Ferreira:2004yd,Barroso:2005sm}: (i)
neutral minimum, (ii) $CP$-violating minimum, (iii) charge-breaking minimum. Here,
neutral minimum breaks only the electroweak symmetry. Whereas, $CP$- and charge-breaking
minima break the respective symmetries, in addition to the electroweak symmetry.
$CP$-violating minimum is phenomenologically
acceptable, however, charge-breaking minimum should be avoided, since violation of
charge symmetry is not found in experiments. Theoretically it is demonstrated that,
in the 2HDM, a neutral minimum do not coexist with either $CP$- or
charge-breaking minima \cite{Ferreira:2004yd,Barroso:2005sm,Ivanov:2006yq,
Ivanov:2007de}. In other words, it is possible to choose a
parameter region of the scalar potential of 2HDM in such a way that the minimum breaks
only the electroweak symmetry spontaneously, and moreover, this can be the global minimum.
This result is appealing and it makes the 2HDM as a viable candidate for the
extension of SM.

The result mentioned above is valid in any model where the scalar sector contains only
two Higgs doublets. Several models are proposed with only the
two Higgs doublets, in order to explain the limitations of SM
\cite{Quigg:2004is,Ellis:2009pz}. One among these
is the scotogenic model \cite{Ma:2006km}, whose motivation is to explain the smallness
of neutrino masses
and the existence of dark matter. This model contains an additional and exact discrete
symmetry $Z_2$, whose purpose is to generate neutrino masses at 1-loop level and also
to accommodate a candidate for dark matter. In order to achieve the motivation of this model,
one of the two Higgs doublets of this model should acquire non-zero VEV and the other
one should acquire zero VEV. We can consider this pattern of VEVs to the Higgs doublets
as the desired vacuum of the scotogenic model. However, by minimizing the scalar
potential of this model,
it is possible for both the Higgs doublets to acquire different patterns of VEVs, and
thus generate different possible vacua. Topics on this subject are discussed in
\cite{Ginzburg:2010wa,Ferreira:2015pfi}
and it is shown that it is possible to make the desired vacuum of the scotogenic model
as the global minimum by restricting the parameter space of the model.
See \cite{Cacciapaglia:2020psm,Rosenlyst:2021tdr}, for alternative proposals
on scotogenic mechanism in composite Higgs models.

In the scotogenic model \cite{Ma:2006km}, neutrinos are Majorana particles.
Since there is no indication
from experiments on the Majorana nature of neutrinos, a priori, it is worth to construct
models
based on Dirac nature of neutrinos. It is for this reason, the scotogenic model has been
modified into Dirac scotogenic model \cite{Farzan:2012sa},
where the neutrinos are purely Dirac particles.
In this later model, three copies of Weyl singlet fermions $\nu^c_\alpha$,
$N_k$, $N^c_k$ are introduced. Here, $\alpha=e,\mu,\tau$, which is the generation
index of lepton family and $k=1,2,3$.
$N_k$ and $N^c_k$ combine to give massive Dirac fermions $N^D_k$.
Whereas, $\nu^c_\alpha$ combine with left-handed neutrinos of the lepton doublets to form
Dirac neutrinos $\nu^D_\alpha$. To forbid Majorana masses for the above Weyl fermions and to
conserve lepton number, an additional and exact symmetry $U(1)_{B-L}$ is proposed.
In the scalar sector of this model, there
exist two complex Higgs doublets ($\Phi$, $\eta$) and a real scalar
singlet ($\chi$). Here, $\Phi$ is the SM-like Higgs doublet.
This model has an additional symmetry $Z_2^{(A)}\times Z_2^{(B)}$,
which prevents masses to Dirac neutrinos at tree level and generate them at 1-loop
level \cite{Farzan:2012sa}. The construction of the model is such that the $Z_2^{(A)}$
symmetry is softly broken but
$Z_2^{(B)}$ is exact symmetry. Hence, the lightest charged particle under
$Z_2^{(B)}$ can be a viable candidate for dark matter.

Like in the case of scotogenic model, in the Dirac scotogenic model as well,
the scalar fields
should acquire VEVs in a specific pattern in order to generate masses for neutrinos
at 1-loop level in a consistent way. This pattern is such that, only $\Phi$ acquires
non-zero VEV, whereas, $\eta$ and $\chi$ acquire zero VEVs \cite{Farzan:2012sa}.
We expect this
pattern of VEVs to constitute a minimum of the model in some parameter region of it.
On the other hand, with the description we have given for the cases of 2HDM
and scotogenic model,
one can expect other possible minima for Dirac scotogenic model, apart from the
desired minimum which is mentioned above. A noteworthy point is that the scalar content
of Dirac scotogenic model is different from that of 2HDM. Hence, the
results we described above for the case of 2HDM need not be applicable to the
Dirac scotogenic model.
More specifically, we may expect some charge-breaking minima to coexist with the
desired minimum of this model. As a result of this, we need to know if the desired
minimum of this model can be made as the global minimum.

In this work, after analyzing the scalar potential, we describe all possible
inequivalent vacua of the Dirac scotogenic
model. We have found that, including the desired minimum of this model, there can
exist eleven different vacua, which includes three charge-breaking minima.
As part of our investigation on global minimum, we have studied
if the desired minimum of this model can coexist with other possible vacua of the model.
In our numerical analysis, we have found that the desired minimum of this model
do not coexist with charge-breaking minima in the viable parameter space of
this model. We have justified this result
by giving an analytical proof to it. On the other hand, the desired
minimum of this model is found to coexist with certain other minima of the model.
In the case that the desired minimum of this model coexist with other minima,
we have given the conditions that need to be satisfied in order to make the desired
minimum of this model as the global minimum. We shown that there exist plenty of
parameter space where the desired minimum of this model is the global minimum.

The study on global minimum of the Dirac scotogenic model will have implications
on the scalar sector of this model, since the analysis is mainly concerned with
the parameters of the scalar potential. One of the phenomenological implications
of the scalar sector of this model is on the signal strength of the Higgs to diphoton
decay $H\to\gamma\gamma$. The signal strength of $H\to\gamma\gamma$, $R_{\gamma\gamma}$,
is measured in the LHC experiment and its value is around one \cite{Workman:2022ynf}.
The additional
contribution to the decay $H\to\gamma\gamma$ in the Dirac scotogenic model
\cite{Farzan:2012sa} is due to the trilinear coupling of the Higgs
with the charged component of $\eta$ field.
As a result of this, the contribution to $R_{\gamma\gamma}$ is determined by the
above trilinear coupling and also by the masses of components of $\eta$ field.
Since the couplings and masses of scalar fields are affected by
the above described analysis of
global minimum, we have studied its implications on
$R_{\gamma\gamma}$. In our analysis, we have found that the experimental value of
$R_{\gamma\gamma}$ can be fitted in this model, irrespective of the fact that the desired
minimum of this model coexist with other minima or not. The fitted value to
$R_{\gamma\gamma}$ in this model, is found to be either less or greater
than one, depending on the parameter choice.

Another implication of the scalar sector of the Dirac scotogenic
model is on the dark matter phenomenology \cite{Workman:2022ynf}.
As described above, in this model, $Z_2^{(B)}$
is an exact symmetry. Hence, the lightest particle charged under the
$Z_2^{(B)}$ can be a candidate for dark matter. We have studied on the possibility of
making any of the additional scalar fields of this model as a candidate
for thermal cold dark matter.

The paper is organized as follows. In the next section, we give a brief description
on the Dirac scotogenic model. In Sec. \ref{s3}, we describe all different
possible minima of this model. In Secs. \ref{s4} and \ref{s5}, we discuss
on making the desired minimum of this model as the global minimum. In Sec.
\ref{s6}, we present our study on the signal strength of Higgs to diphoton decay.
In Sec. \ref{s7}, we discuss on the possibility of a scalar dark matter
candidate in this model. In Sec. \ref{sadd}, we have compared the phenomenology
of the Dirac scotogenic model with that of scotogenic model.
We present the conclusions of our work in Sec. \ref{s8}.
In appendix \ref{app}, we describe analytical arguments in order to
justify some of the numerical results of Sec. \ref{s5}.

\section{Dirac scotogenic model}
\label{s2}

We have given a brief introduction to the Dirac scotogenic model \cite{Farzan:2012sa}
in Sec. \ref{s1}. In this work, we follow the original model of this, which is proposed
in \cite{Farzan:2012sa}. Apart from this, other models are also proposed which
have the idea of scotogenic masses to Dirac neutrinos \cite{Ma:2016mwh,Wang:2017mcy,
Ma:2019yfo,Leite:2020wjl,Guo:2020qin,Bernal:2021ezl,Borah:2022phw}. In the Dirac
scotogenic model \cite{Farzan:2012sa}, additional scalars and fermionic fields are
introduced along with the additional symmetry $U(1)_{B-L}\times Z_2^{(A)}\times Z_2^{(B)}$.
The field content of this model, which is relevant to lepton sector, and their charge
assignments are given in Tab. \ref{t1}.
\begin{table}[!h]
\centering
\begin{tabular}{|c|c|c|c|c|c|}\hline
Field & $SU(2)_L$ & $U(1)_Y$ & $U(1)_{B-L}$ & $Z_2^{(A)}$ & $Z_2^{(B)}$ \\\hline
$L_\alpha=(\nu_\alpha,\ell_\alpha)$ & $2$ & $-1/2$ & $-1$ & $+$ & $+$ \\
$\ell^c_\alpha$ & $1$ & $1$ & $1$ & $+$ & $+$ \\
$\nu^c_\alpha$ & $1$ & $0$ & $1$ & $-$ & $+$ \\ \hline
$\Phi^T=(\phi^+,\phi^0)$ & $2$ & $1/2$ & $0$ & $+$ & $+$ \\
$\eta^T=(\eta^+,\eta^0)$ & $2$ & $1/2$ & $0$ & $+$ & $-$ \\
$\chi$ & $1$ & $0$ & $0$ & $-$ & $-$ \\ \hline
$N_k$ & $1$ & $0$ & $-1$ & $+$ & $-$ \\
$N_k^c$ & $1$ & $0$ & $1$ & $+$ & $-$ \\ \hline
\end{tabular}
\caption{Fields in the lepton sector of Dirac scotogenic model
\cite{Farzan:2012sa} along with their charge assignments. As described
in Sec. \ref{s1}, $\Phi,\eta$ and $\chi$ are scalars. Rest of the fields are fermionic.}
\label{t1}
\end{table}
Here, the symmetry $U(1)_{B-L}$ can be either global or gauged \cite{Farzan:2012sa}.
In this work, we have taken this to be global and it is an exact symmetry.
With the charge assignments of Tab. \ref{t1}, the allowed interaction terms
in the Lagrangian are
\begin{equation}
-{\cal L}=y_{\alpha\beta}(\nu_\alpha\phi^{+^*}+\ell_\alpha\phi^{0^*})\ell^c_\beta
+f_{\alpha k}(\nu_\alpha\eta^0-\ell_\alpha\eta^+)N^c_k
+h_{k\alpha}N_k\nu^c_\alpha\chi+m_{N_k}N_kN^c_k+h.c.
\label{lag}
\end{equation}
Here, $m_{N_k}$ is the Dirac mass for $N^D_k=(N_k,N^c_k)$. The couplings $y_{\alpha\beta}$
in Eq. (\ref{lag}) generate masses to charged leptons
after the electroweak symmetry breaking, where $\Phi$ acquires non-zero VEV.
On the other hand, the terms of Eq. (\ref{lag}) do not generate masses to
light neutrinos at tree level,
since $\eta$ and $\chi$ acquire zero VEVs, which is due to the fact that both
these fields are charged under the exact symmetry $Z_2^{(B)}$. However, if there
exist a trilinear term among $\Phi$, $\eta$ and $\chi$, neutrinos acquire
masses at 1-loop level in this model. Shortly below, we explain how such a trilinear
term can be present in this model. We notice from Tab. \ref{t1} that
all the additional fields of this model are charged under the $Z_2^{(B)}$
symmetry. Since this symmetry is exact, the lightest among these additional
fields can be a candidate for dark matter. Later in this work, we discuss on
the possibility of a scalar dark matter in this model.

As discussed in the previous section, the Dirac scotogenic model is motivated
from the scotogenic model \cite{Ma:2006km}, where neutrinos are Majorana
particles. We notice that there is an analogy between these two models in terms
of field content. Instead of $(N_k,N^c_k)$, there exist $N^M_k$ in the scotogenic
model, which is a Majorana field. The scalar sector of the scotogenic model
is same as that of Dirac scotogenic model, except for the singlet field $\chi$.
As a result of this, the third term of Eq. (\ref{lag})
does not exist in the Lagrangian of scotogenic model. Moreover, by replacing
$(N_k,N^c_k)$ with a Majorana $N^M_k$ field in Eq. (\ref{lag}), we get
corresponding terms in the scotogenic model \cite{Ma:2006km}.
Later in this work, we compare the above two models
in terms of neutrino masses and other physically observable quantities.

The scalar potential of the Dirac scotogenic model is \cite{Farzan:2012sa}
\begin{eqnarray}
V&=&\mu_1^2\Phi^\dagger\Phi+\mu_2^2\eta^\dagger\eta+\frac{1}{2}\mu_3^2\chi^2+\frac{1}{2}
\lambda_1(\Phi^\dagger\Phi)^2+\frac{1}{2}\lambda_2(\eta^\dagger\eta)^2+\lambda_3
(\Phi^\dagger\Phi)(\eta^\dagger\eta)
\nonumber \\
&&+\lambda_4(\Phi^\dagger\eta)(\eta^\dagger\Phi)
+\frac{1}{2}[\lambda_5(\Phi^\dagger\eta)^2+h.c.]+\frac{1}{4}\lambda_6\chi^4+\frac{1}{2}
\lambda_7(\Phi^\dagger\Phi)\chi^2+\frac{1}{2}\lambda_8(\eta^\dagger\eta)\chi^2
\nonumber \\
&&+A\chi[\Phi^\dagger\eta+h.c.]
\label{pot}
\end{eqnarray}
Here, we have chosen the parameter $A$ to be real by fixing the phases in $\Phi$ and
$\eta$. For this particular choice, the parameter $\lambda_5$ can in general be
complex. However, in the analysis of \cite{Farzan:2012sa}, $\lambda_5$ is taken
to be real for the sake of simplicity. We discuss about $\lambda_5$ in the context
our work, later in the next section. The terms of Eq. (\ref{pot}) generate masses
to physical fields of this model, after the electroweak symmetry breaking, where
$\Phi$ acquires non-zero VEV and the $\eta,\chi$ acquire zero VEVs. The non-zero
VEV for $\Phi$ can be taken as $\langle\phi^0\rangle=v_{EW}=174$ GeV, which
is the electroweak symmetry breaking scale. We see that,
after this symmetry breaking, $\phi^+$ and imaginary part of $\phi^0$
become Goldstone bosons. The real part of $\phi^0$ is physical and we
identify $Re(\phi^0)=H$ as the Higgs boson. The charged component of $\eta$
is physical. On the other hand, the neutral component of $\eta$ has mixing with
$\chi$ through the last term of Eq. (\ref{pot}). As a result of this, we
write $\eta^0=(\eta_R^0+i\eta_I^0)/\sqrt{2}$. Now, the mass spectrum of the
physical scalar fields in the Dirac scotogenic model is
\cite{Farzan:2012sa}
\begin{eqnarray}
&&m_H^2=2\lambda_1v_{EW}^2,
\nonumber \\
&&m_{\eta^+}^2=\mu_2^2+\lambda_3v_{EW}^2,
\nonumber \\
&&m_{\eta_I^0}^2\equiv m_{\zeta_3}^2=\mu_2^2+(\lambda_3+\lambda_4-\lambda_5)v_{EW}^2,
\nonumber \\
&&M_{\eta_R^0,\chi}^2=\left(\begin{array}{cc}
\mu_2^2+(\lambda_3+\lambda_4+\lambda_5)v_{EW}^2 & \sqrt{2}Av_{EW} \\
\sqrt{2}Av_{EW} & \mu_3^2+\lambda_7v_{EW}^2 \end{array}\right)
\label{mas-ds}
\end{eqnarray}
Here, $M_{\eta_R^0,\chi}^2$ gives the mixing masses between $\eta_R^0$ and
$\chi$, whose mass eigenstates are denoted by $\zeta_{1,2}$. Also here,
for the sake of notational simplicity, we have written $\eta_I^0=\zeta_3$.

The last term of Eq. (\ref{pot}) breaks the $Z_2^{(A)}$ symmetry softly.
This term is necessary in order to generate masses to neutrinos at 1-loop level
\cite{Farzan:2012sa}, and moreover, this is the trilinear term that we have discussed
above. Now, with the $A$-term of Eq. (\ref{pot}) and with the interaction terms
of Eq. (\ref{lag}), neutrinos acquire masses at 1-loop, whose expressions are given
below \cite{Farzan:2012sa}.
\begin{eqnarray}
({\cal M}_\nu)_{\alpha\beta}&=&(f\Lambda h)_{\alpha\beta}=
\sum_{K=1}^3f_{\alpha k}\Lambda_k h_{k\beta},
\nonumber \\
\Lambda_k&=&\frac{\sin(2\theta)}{32\pi^2\sqrt{2}}
m_{N_k}\left[\frac{m_{\zeta_1}^2}{m_{\zeta_1}^2-m_{N_k}^2}
\ln\frac{m_{\zeta_1}^2}{m_{N_k}^2}-\frac{m_{\zeta_2}^2}{m_{\zeta_2}^2-m_{N_k}^2}
\ln\frac{m_{\zeta_2}^2}{m_{N_k}^2}\right]
\label{mnu}
\end{eqnarray}
Here, $\theta$ is the mixing angle between $\eta^0_R$ and $\chi$. As a
result of this mixing, the diagonal masses for these fields are denoted by
$m_{\zeta_1}$ and $m_{\zeta_2}$. As already described before, this mixing is
arising due to the $A$-term
of Eq. (\ref{pot}). As a result of this, $\theta$ is proportional to the $A$ parameter.
Since the $A$-term breaks $Z_2^{(A)}$ symmetry softly, the parameter $A$, and hence,
the $\theta$ can be small. The small mass for neutrinos in Eq. (\ref{mnu})
can be explained either through small mixing angle $\theta$, or by large value for $m_{N_k}$,
or by taking degenerate masses for $\zeta_{1,2}$,
apart from the loop suppression factor. It is to be noted that the masses of
$\zeta_{1,2}$ depend on various parameters, which are given in Eq. (\ref{mas-ds}).
It is possible to fine tune these parameters in such way that $m_{\zeta_1}$ and
$m_{\zeta_2}$ are nearly degenerate, which gives an additional suppression for
neutrino masses in Eq. (\ref{mnu}).

Apart from the smallness of neutrino masses, we need to explain the observed neutrino
mixing angles \cite{Workman:2022ynf} in the Dirac scotogenic model. In order to
explain this, we use the Casas-Ibarra parametrization \cite{Casas:2001sr}, and
thereby, the Yukawa couplings in Eq. (\ref{mnu}) can be expressed as
\begin{eqnarray}
f&=&U\sqrt{m_\nu}R\sqrt{\Lambda}^{-1},\quad
h=\sqrt{\Lambda}^{-1}S^\dagger\sqrt{m_\nu}V^\dagger,
\nonumber \\
m_\nu&=&{\rm diag}(m_{\nu_1},m_{\nu_2},m_{\nu_3})
\label{fh}
\end{eqnarray}
Here, $m_{\nu_i}$, where $i=1,2,3$, are the neutrino mass eigenvalues.
$R$ and $S$ are in general
complex matrices, which satisfy $RS^\dagger=I$. After using the above
parametrizations for $f$ and $h$ in Eq. (\ref{mnu}), we get
$U^\dagger{\cal M}_\nu V=m_\nu$, which is the desired relation for
diagonalizing the ${\cal M}_\nu$. Here, $U=U_{PMNS}$ is identified as the
Pontecorvo-Maki-Nakagawa-Sakata matrix, which is parametrized in terms of the three
neutrino mixing angles and a $CP$ violating Dirac phase \cite{Workman:2022ynf}.
$V$ is a unitary matrix which rotate the right-handed neutrino fields from
flavor to mass eigenstates. As a result of the above given description, the
neutrino mixing angles in the Dirac scotogenic model can be explained by
parametrizing the Yukawa couplings as in Eq. (\ref{fh}). In the parametrizations
of $f$ and $h$, the neutrino
mass eigenvalues can be chosen either in normal or inverted ordering, in order
to fit the solar and atmospheric mass-square differences of neutrinos
\cite{Workman:2022ynf}.

The expression for neutrino masses in the Dirac scotogenic model, which is given
in Eq. (\ref{mnu}) is similar to the corresponding
expression of the scotogenic model \cite{Ma:2006km}.
In the context of neutrino mass generation, the difference
between the above two models is described below. The neutrino masses in the
Dirac scotogenic model are driven by two different Yukawa couplings, whereas,
in the scotogenic model these masses are driven by one kind of Yukawa couplings.
The parametrizations of Yukawa couplings in the Dirac scotogenic model, which
are given in Eq. (\ref{fh}), are similar to that in the scotogenic model, which
can be seen from \cite{Hundi:2022iva}. In \cite{Hundi:2022iva}, we have worked on the
lepton flavor violating (LFV) decays of $Z$ and Higgs boson in the scotogenic
model, where we have also done numerical analysis on fitting the neutrino masses
and mixing angles in this model. This numerical analysis can be analogously worked
in the Dirac scotogenic model.

To test the Dirac scotogenic model in collider experiments, such as the LHC,
we should take the masses of all additional fields to be around few hundred
GeV. Now, using the discussion given before, for $m_{N_K}\sim$ 1 TeV,
the small masses for neutrinos can be
explained if either of the following quantities are taken to be small:
$\theta$ or $m_{\zeta_1}-m_{\zeta_2}$. In the limit that these
quantities are small, we see that $\Lambda_k$ of Eq. (\ref{mnu}) becomes small.
Hence, from Eq. (\ref{fh}), the Yukawa couplings $f$ and $h$ can become
${\cal O}(1)$ in some region of parameter space. The couplings $f_{\alpha k}$
drive LFV processes in this model which can have significant branching ratios, since
$f_{\alpha k}$ are not suppressed and $m_{N_K}\sim$ 1 TeV.
Later in this work, we describe about LFV processes of this model.

As described before, in the Dirac scotogenic model, $\Phi$ acquires non-zero
VEV and $\eta,\chi$ should acquire zero VEVs. This pattern of VEVs can be achieved
by minimizing the potential of Eq. (\ref{pot}) for $\mu_1^2<0$. Here we notice
that in some parameter region of the scalar potential the above pattern of VEVs
constitute a minimum for the model. By choosing different parameter regions of
the scalar potential, it is possible to find other minima of this model.
In the next section, we argue that the above minimum required for the
Dirac scotogenic model is only one possible minima of this model.

\section{Possible minima of the Dirac scotogenic model}
\label{s3}

In this section, we describe different possible minima of the Dirac scotogenic model,
after analyzing the scalar potential of it. The scalar potential of this model
is described in Eq. (\ref{pot}), which consists of the fields
$\Phi$, $\eta$ and $\chi$. At the minimum of the scalar potential,
either of these scalar fields can acquire non-zero VEVs. First, let us consider the case
where $\langle\Phi\rangle\neq0$ and $\langle\eta\rangle=0$. In this case, using
the $SU(2)$ transformation, the VEV of $\Phi$ can always be brought into a form where
the neutral component of it acquires non-zero and real VEV.
As a result of this, there can exist
two different minima, depending on whether $\chi$ acquires a VEV or not.
These possible minima are
\begin{eqnarray}
&&N1:\quad\langle\Phi\rangle=\left(\begin{array}{c}0\\v_{EW}\end{array}\right),\quad
\langle\eta\rangle=\left(\begin{array}{c}0\\0\end{array}\right),\quad\langle\chi\rangle=0.
\label{n1}
\\
&&N2:\quad\langle\Phi\rangle=\left(\begin{array}{c}0\\v_{\phi(2)}\end{array}\right)\quad
\langle\eta\rangle=
\left(\begin{array}{c}0\\0\end{array}\right),\quad\langle\chi\rangle=v_{\chi(2)},
\label{n2}
\end{eqnarray}
where $v_{\phi(2)}$ and
$v_{\chi(2)}$ are some non-zero real variables. Here, $N1,N2$ are neutral minima
which break only the electroweak symmetry. $N1$ is the desired minimum of the Dirac
scotogenic model, which we have considered to be the true model of our universe.
Hence, we have equated the VEV of $\Phi$ field to $v_{EW}$. On the other hand, $N2$
is one possible minima of the scalar potential of the Dirac scotogenic model. But
otherwise, this minimum is not consistent with the model framework. Hence, $N2$ does not
represent the observable world of our universe. As a result of this, the VEVs of
$\Phi$ and $\chi$ in this minimum are some variables and need not represent the
electroweak symmetry breaking scale. Now, let us consider the case where
$\langle\Phi\rangle=0$ and $\langle\eta\rangle\neq0$. In this case, in analogy to the
description given above, the following two minima are possible:
\begin{eqnarray}
&&N3:\quad\langle\Phi\rangle=\left(\begin{array}{c}0\\0\end{array}\right),\quad
\langle\eta\rangle=\left(\begin{array}{c}0\\v_{\eta(3)}\end{array}\right),
\quad\langle\chi\rangle=0.
\label{n3}
\\
&&N4:\quad\langle\Phi\rangle=\left(\begin{array}{c}0\\0\end{array}\right)\quad
\langle\eta\rangle=\left(\begin{array}{c}0\\v_{\eta(4)}\end{array}\right),
\quad\langle\chi\rangle=v_{\chi(4)}.
\label{n4}
\end{eqnarray}
The non-zero entries in the above equations are real and arbitrary. The minima
$N3$ and $N4$ are unphysical, since they do not generate masses to SM fermions
of this model.

In the case where $\langle\Phi\rangle=0=\langle\eta\rangle$, there can exist one
non-trivial minimum, which is given below.
\begin{equation}
N5:\quad\langle\Phi\rangle=\left(\begin{array}{c}0\\0\end{array}\right)\quad
\langle\eta\rangle=\left(\begin{array}{c}0\\0\end{array}\right),
\quad\langle\chi\rangle=v_{\chi(5)}.
\label{n5}
\end{equation}
Here, the VEV of $\chi$ is real and non-zero variable. The minimum $N5$ is clearly
unphysical, since it doesn't break the electroweak symmetry.
Finally, in the case where $\langle\Phi\rangle\neq0\neq\langle\eta\rangle$, the
following six minima are possible:
\begin{eqnarray}
&&N6:\quad\langle\Phi\rangle=\left(\begin{array}{c}0\\v_{\phi(6)}\end{array}\right),\quad
\langle\eta\rangle=\left(\begin{array}{c}0\\v_{\eta(6)}\end{array}\right),
\quad\langle\chi\rangle=0.
\label{n6}
\\
&&N7:\quad\langle\Phi\rangle=\left(\begin{array}{c}0\\v_{\phi(7)}\end{array}\right),\quad
\langle\eta\rangle=\left(\begin{array}{c}0\\iv_{\eta(7)}\end{array}\right),
\quad\langle\chi\rangle=0.
\label{n7}
\\
&&N8:\quad\langle\Phi\rangle=\left(\begin{array}{c}0\\v_{\phi(8)}\end{array}\right)\quad
\langle\eta\rangle=\left(\begin{array}{c}0\\v_{\eta(8)}\end{array}\right),
\quad\langle\chi\rangle=v_{\chi(8)}.
\label{n8}
\\
&&C9:\quad\langle\Phi\rangle=\left(\begin{array}{c}0\\v_{\phi(9)}\end{array}\right),\quad
\langle\eta\rangle=\left(\begin{array}{c}c_{\eta(9)}\\v_{\eta(9)}\end{array}\right),
\quad\langle\chi\rangle=0.
\label{c9}
\\
&&C10:\quad\langle\Phi\rangle=\left(\begin{array}{c}0\\v_{\phi(10)}\end{array}\right)\quad
\langle\eta\rangle=\left(\begin{array}{c}c_{\eta(10)}\\iv_{\eta(10)}\end{array}\right),
\quad\langle\chi\rangle=0.
\label{c10}
\\
&&C11:\quad\langle\Phi\rangle=\left(\begin{array}{c}0\\v_{\phi(11)}\end{array}\right)\quad
\langle\eta\rangle=\left(\begin{array}{c}c_{\eta(11)}\\v_{\eta(11)}\end{array}\right),
\quad\langle\chi\rangle=v_{\chi(11)}.
\label{c11}
\end{eqnarray}
Here, $C9$, $C10$ and $C11$ are charge-breaking minima, where the charged component of
$\eta$ acquires non-zero VEV. On the other hand, $N6$, $N7$ and $N8$ break only the
electroweak symmetry. In obtaining the forms of VEVs in Eqs. (\ref{n6})$-$(\ref{c11}), we
have used the $SU(2)$ transformation \cite{Branco:2011iw} on $\Phi$ and $\eta$.
It is to be noticed that the variables of the form $v_{\phi(k)},v_{\eta(k)},c_{\eta(k)},
v_{\chi(k)}$ in Eqs. (\ref{n6})$-$(\ref{c11}) are real and non-zero. While obtaining
that these variables are real, we have taken the $\lambda_5$ parameter of
Eq. (\ref{pot}) to be real. Shortly below, we give a demonstration about this.
We notice that the VEV of $\eta$ in Eqs. (\ref{n7}) and (\ref{c10}) is complex,
which tells something about $CP$ symmetry in the minima of $N7$ and $C10$.
As we have taken $\lambda_5$ to be real, the
minimum $N7$ respect the $CP$ symmetry, since Eqs. (\ref{pot}) and (\ref{n7}) are
invariant under the following $CP$ transformation: $\Phi\to\Phi^*,\eta\to-\eta^*,
\chi\to-\chi$. On the other hand, $C10$ breaks the $CP$ symmetry spontaneously,
apart from the charge and electroweak symmetries.

After using the $SU(2)$ transformation \cite{Branco:2011iw}, the general
structure of the VEVs of the scalar fields in the Dirac scotogenic model can be written as
\begin{equation}
\langle\Phi\rangle=\left(\begin{array}{c}0\\v_{\phi}\end{array}\right)\quad
\langle\eta\rangle=\left(\begin{array}{c}c_{\eta}\\v_{\eta}\end{array}\right),
\quad\langle\chi\rangle=v_{\chi}
\label{vev}
\end{equation}
In the above equation, $v_\phi,c_\eta,v_\chi$ are real variables and
$v_\eta$ is in general a complex variable. Apart from $v_\eta$, the other
complex variable in this model is the $\lambda_5$ parameter of Eq. (\ref{pot}).
As a result of this, we write the forms for $\lambda_5$ and $v_{\eta}$ as
\begin{equation}
\lambda_5=|\lambda_5|e^{i\theta_5},\quad
v_{\eta}=|v_{\eta}|e^{i\theta_\eta}
\end{equation}
Here, $\theta_5$ and $\theta_\eta$ are phases in $\lambda_5$ and $v_{\eta}$,
respectively. After plugging Eq. (\ref{vev}) in Eq. (\ref{pot}), the relevant part
of the potential is
\begin{equation}
\langle V\rangle=|\lambda_5|v_{\phi}^2|v_{\eta}|^2\cos(\theta_5+2\theta_\eta)
+2Av_{\chi}v_{\phi}|v_{\eta}|\cos\theta_\eta
\label{minv}
\end{equation}
In the above equation, we have written only that part of the potential which contains
only the phases $\theta_5$ and $\theta_\eta$. The extremum for $\langle V\rangle$
with respect to these phases is found to be
\begin{equation}
\theta_5=(m-2n)\pi,\quad\theta_\eta=n\pi
\label{ext}
\end{equation}
Here, $m,n$ are any integers. Using the above relations, we see that
$\lambda_5$ and $v_{\eta}$ are real quantities at any extremum of the potential.
To see if the values of Eq. (\ref{ext}) correspond to
the minimum of Eq. (\ref{minv}), we need to evaluate the second order derivatives
of $\langle V\rangle$ with respect to $\theta_5$ and $\theta_\eta$. As a result of this,
we get the following matrix
\begin{equation}
M_V=\left.\left(\begin{array}{cc}
\frac{\partial^2\langle V\rangle}{\partial\theta_5^2} &
\frac{\partial^2\langle V\rangle}{\partial\theta_5\partial\theta_\eta} \\
\frac{\partial^2\langle V\rangle}{\partial\theta_5\partial\theta_\eta} &
\frac{\partial^2\langle V\rangle}{\partial\theta_\eta^2}
\end{array}\right)\right|_{\theta_5=(m-2n)\pi,\theta_\eta=n\pi}
\end{equation}
After demanding that the eigenvalues of $M_V$ are positive, and since
$\lambda_5$ and $v_{\eta}$ are real at the minimum of the potential, we get the
following conditions
\begin{equation}
\lambda_5<0,\quad Av_{\phi}v_{\eta}v_{\chi}<0
\label{cond}
\end{equation}

Using the analysis, which is described in the previous paragraph, we notice
that in the minima of Eqs. (\ref{n8}) and (\ref{c11}), the quantities $v_{\eta(8)}$
and $v_{\eta(11)}$ should be real. Moreover, in these minima, $\lambda_5$ should be
a real parameter and the conditions of Eq. (\ref{cond}) should be satisfied.
Now, in the analysis of previous paragraph, let us consider the case where
$v_\chi=0$. In this case, at the minimum of the potential, the individual phases
in $\lambda_5$ and $v_\eta$ cannot be determined. However, if we choose $\lambda_5$
to be real, then $v_\eta$ can be either real or purely imaginary at the minimum
of the potential. As a result of this, the VEV of the neutral component of $\eta$
is taken to be real in Eqs. (\ref{n6}) and (\ref{c9}), whereas, this quantity
is taken to be purely imaginary in Eqs. (\ref{n7}) and (\ref{c10}). Let us mention
here that $\lambda_5$ can in general be complex in the minima of
Eqs. (\ref{n1})$-$(\ref{n5}). However, to simplify our numerical analysis, we
have taken $\lambda_5$ to be real in the rest of this work.

We have described that the VEV structures given in Eqs. (\ref{n1})$-$(\ref{c11})
as possible minima of the model. In order to clarify this point, we refer each of these VEV
structures as a stationary point (SP). An SP becomes a minimum if the
following two conditions are satisfied: (1) minimization conditions of the
scalar potential, (2) mass-square eigenvalues
of scalar fields, which are not Goldstone bosons, are positive.
The minimization conditions, which should be
satisfied by the SPs of Eqs. (\ref{n1})$-$(\ref{c11}), are given in
Tab. \ref{t2}.
\begin{table}[!h]
\centering
\begin{tabular}{|c|l|} \hline
$N1$ & $\mu_1^2+\lambda_1v_{EW}^2=0$ \\ \hline
$N2$ & $A=0,\quad \mu_1^2+\lambda_1v_{\phi(2)}^2+\frac{1}{2}\lambda_7v_{\chi(2)}^2=0,
\quad \mu_3^2+\lambda_6v_{\chi(2)}^2+\lambda_7v_{\phi(2)}^2=0$ \\ \hline
$N3$ & $\mu_2^2+\lambda_2v_{\eta(3)}^2=0$ \\ \hline
$N4$ & $A=0,\quad \mu_2^2+\lambda_2v_{\eta(4)}^2+\frac{1}{2}\lambda_8v_{\chi(4)}^2=0,
\quad \mu_3^2+\lambda_6v_{\chi(4)}^2+\lambda_8v_{\eta(4)}^2=0$ \\ \hline
$N5$ & $\mu_3^2+\lambda_6v_{\chi(5)}^2=0$ \\ \hline
$N6$ & $A=0,\quad\mu_1^2+\lambda_1v_{\phi(6)}^2+(\lambda_3+\lambda_4+\lambda_5)
v_{\eta(6)}^2=0$,\\
     & $\mu_2^2+\lambda_2v_{\eta(6)}^2+(\lambda_3+\lambda_4+\lambda_5)
v_{\phi(6)}^2=0$ \\ \hline
$N7$ & $\mu_1^2+\lambda_1v_{\phi(7)}^2+(\lambda_3+\lambda_4-\lambda_5)
v_{\eta(7)}^2=0,\quad \mu_2^2+\lambda_2v_{\eta(7)}^2+(\lambda_3+\lambda_4-\lambda_5)
v_{\phi(7)}^2=0$ \\ \hline
$N8$ & $\mu_1^2+\lambda_1v_{\phi(8)}^2+(\lambda_3+\lambda_4+\lambda_5)
v_{\eta(8)}^2+\frac{1}{2}\lambda_7v_{\chi(8)}^2+Av_{\chi(8)}v_{\eta(8)}/v_{\phi(8)}=0,$ \\
     & $\mu_2^2+\lambda_2v_{\eta(8)}^2+(\lambda_3+\lambda_4+\lambda_5)v_{\phi(8)}^2
+\frac{1}{2}\lambda_8v_{\chi(8)}^2+Av_{\chi(8)}v_{\phi(8)}/v_{\eta(8)}=0,$ \\
     & $\mu_3^2+\lambda_6v_{\chi(8)}^2+\lambda_7
v_{\phi(8)}^2+\lambda_8v_{\eta(8)}^2+2Av_{\phi(8)}v_{\eta(8)}/v_{\chi(8)}=0$ \\ \hline
$C9$ & $A=0,\quad \lambda_4+\lambda_5=0,\quad
\mu_1^2+\lambda_1v_{\phi(9)}^2+\lambda_3(c_{\eta(9)}^2+v_{\eta(9)}^2)=0,$ \\
     & $\mu_2^2+\lambda_2(c_{\eta(9)}^2+v_{\eta(9)}^2)+\lambda_3v_{\phi(9)}^2=0$ \\ \hline
$C10$ & $\lambda_4-\lambda_5=0,\quad
\mu_1^2+\lambda_1v_{\phi(10)}^2+\lambda_3(c_{\eta(10)}^2+v_{\eta(10)}^2)=0,$ \\
     & $\mu_2^2+\lambda_2(c_{\eta(10)}^2+v_{\eta(10)}^2)+\lambda_3v_{\phi(10)}^2=0$ \\ \hline
$C11$ & $Av_{\chi(11)}+(\lambda_4+\lambda_5)v_{\eta(11)}v_{\phi(11)}=0,$ \\
 & $\mu_1^2+\lambda_1v_{\phi(11)}^2+\lambda_3(c_{\eta(10)}^2+
v_{\eta(11)}^2)+\frac{1}{2}\lambda_7v_{\chi(11)}^2=0,$
\\
 & $\mu_2^2+\lambda_2(c_{\eta(11)}^2+v_{\eta(11)}^2)+\lambda_3v_{\phi(11)}^2
+\frac{1}{2}\lambda_8v_{\chi(11)}^2=0,$ \\
 & $\mu_3^2+\lambda_6v_{\chi(11)}^2+\lambda_7v_{\phi(11)}^2
+\lambda_8(c_{\eta(11)}^2+v_{\eta(11)}^2)+2Av_{\phi(11)}v_{\eta(11)}/v_{\chi(11)}=0$ \\ \hline
\end{tabular}
\caption{Minimization conditions which should be satisfied at the SPs of
Eqs. (\ref{n1})$-$(\ref{c11}).}
\label{t2}
\end{table}
Shortly below, we describe on how we compute the mass-square eigenvalues of scalar
fields at the SPs of Eqs. (\ref{n1})$-$(\ref{c11}). It is to be noted that,
in the Dirac scotogenic model, the scalar sector consists
of nine real degrees of freedom. Out of these nine, some of them may become Goldstone
bosons, since the electroweak and charge symmetries are spontaneously broken by
some of the SPs of
Eqs. (\ref{n1})$-$(\ref{c11}). The SP $N5$ of Eq. (\ref{n5}) does not break
either of these symmetries, and hence, in this case all the nine scalar fields become
massive.

As stated before, the general structure of VEVs of scalar fields in
the Dirac scotogenic model is given by Eq. (\ref{vev}). Hence,
in order to compute the mass eigenstates, we parametrize these fields as
\begin{equation}
\Phi=\left(\begin{array}{c} (\phi_R^1+i\phi_I^1)/\sqrt{2} \\
v_\phi+(\phi_R^0+i\phi_I^0)/\sqrt{2} \end{array}\right),\quad
\eta=\left(\begin{array}{c} c_\eta+(\eta_R^1+i\eta_I^1)/\sqrt{2} \\
v_\eta+(\eta_R^0+i\eta_I^0)/\sqrt{2} \end{array}\right),\quad
\chi=v_\chi+\chi_R
\label{para}
\end{equation}
Now, in the case of Eq. (\ref{n1}), which corresponds to the desired minimum of the Dirac
scotogenic model, $\phi_R^1,\phi_I^1,\phi_I^0$ become Goldstone bosons. Moreover,
for this minimum, we have $v_\phi=v_{EW}$ and $c_\eta=v_\eta=v_\chi=0$. Now, after
identify $\phi_R^0=H$ as the Higgs boson and $\eta^+=(\eta_R^1+i\eta_I^1)/\sqrt{2}$,
we see that the mass spectrum of physical fields in the case of $N1$ matches with
that of Eq. (\ref{mas-ds}).
Similar to what we described above, the mass spectrum of scalar
fields at other SPs
of Eqs. (\ref{n2})$-$(\ref{c11}) are computed accordingly. It should
be noted here that, even if any of these SPs become minima, the scalar fields
in these minima are unphysical, since these minima
do not represent the physical world of ours. In our
analysis, which we will present later, we have computed the scalar masses at
the SPs of Eqs. (\ref{n2})$-$(\ref{c11}) numerically. We have found in our analysis
that each of the SPs given in Eqs. (\ref{n1})$-$(\ref{c11}) can
become a minimum in some region of parameter space of the scalar potential.
Since multiple minima can exist for the scalar potential of the
Dirac scotogenic model, we need to know if $N1$, which is the desired minimum
of this model, can become the global minimum. The next few sections discuss
about this.

\section{Relative depths in potential}
\label{s4}

At the end of the last section, we have noted that $N1$ is one possible minimum
among other minima of the Dirac scotogenic model. In order to address if $N1$ is
the global minimum of this model, we need to know if $N1$ coexist with other minima.
If it coexist with other minima, we demand that the potential depth at $N1$ is
lower as compared to that at other minima, so that $N1$ is the global minimum
of this model. In this section,
we assume that the minimum $N1$ to coexist with other minima in some region of
parameter space. We then calculate the differences in the depth of potential at
$N1$ and at other minima. Using these quantities, we predict on the possibility of
making $N1$ as the global minimum of this model. The results obtained in this
section are helpful for the next section, where we study on the coexistence of $N1$
with other minima.

In order to compute the difference between the
value of potential at $N1$ and at any other SP, we
follow the work of \cite{Ferreira:2004yd,Barroso:2005sm,Ferreira:2019hfk},
which is based on a formalism of bilinears. For some works using the bilinear formalism,
see \cite{Velhinho:1994np,Nishi:2006tg,Maniatis:2006fs,
Ivanov:2014doa,Ferreira:2016tcu}. Using this formalism, we notice that,
except for the last term of Eq. (\ref{pot}), other terms of the scalar potential
are either quadratic or quartic. As a result of this, we define the following
bilinears:
\begin{equation}
x_1=\Phi^\dagger\Phi,\quad x_2=\eta^\dagger\eta,\quad x_3=\chi^2,\quad
x_4=Re(\Phi^\dagger\eta),\quad x_5=Im(\Phi^\dagger\eta)
\label{bi}
\end{equation}
Apart from these, we also define the following matrices:
\begin{equation}
X=\left(\begin{array}{c}x_1\\x_2\\x_3\\x_4\\x_5\end{array}\right),\quad
M_2=\left(\begin{array}{c}\mu_1^2\\\mu_2^2\\\frac{1}{2}\mu_3^2\\0\\0\end{array}\right),\quad
M_4=\left(\begin{array}{ccccc}
\lambda_1 & \lambda_3 & \frac{1}{2}\lambda_7 & 0 & 0 \\
\lambda_3 & \lambda_2 & \frac{1}{2}\lambda_8 & 0 & 0 \\
\frac{1}{2}\lambda_7 & \frac{1}{2}\lambda_8 & \frac{1}{2}\lambda_6 & 0 & 0 \\
0 & 0 & 0 & 2(\lambda_4+\lambda_5) & 0 \\
0 & 0 & 0 & 0 & 2(\lambda_4-\lambda_5) \\
\end{array}\right)
\label{mat}
\end{equation}
Now, using Eqs. (\ref{bi}) and (\ref{mat}), the scalar potential of Eq. (\ref{pot})
can be expressed as
\begin{eqnarray}
&&V=V_2+V_3+V_4,
\nonumber \\
&&V_2=M_2^TX,\quad V_3=2A\chi x_4,\quad V_4=\frac{1}{2}X^TM_4X
\end{eqnarray}
At any SP of Eqs. (\ref{n1})$-$(\ref{c11}), the scalar potential has to satisfy
the minimization conditions. Hence, we get the following relation \cite{Ferreira:2019hfk}
\begin{equation}
\sum_i\varphi_i\left.\frac{\partial V}{\partial\varphi_i}\right|_{SP}=0\implies
2(V_2)_{SP}+3(V_3)_{SP}+4(V_4)_{SP}=0
\end{equation}
In the above equation, $\varphi_i$ represent any real scalar degree of freedom of the
Dirac scotogenic model. Here, $(V_i)_{SP}$, where $i=2,3,4$, is the value of
$V_i$ evaluated at an SP. Using the above relation, the value of scalar potential
evaluated at an SP is found to be
\begin{equation}
V_{SP}=\frac{1}{2}(V_2)_{SP}+\frac{1}{4}(V_3)_{SP}
\end{equation}

We define $X_{SP}$ as the matrix $X$ evaluated at an SP. We then
define the following quantity at any SP:
\begin{equation}
V^\prime_{SP}=M_2+M_4X_{SP}
\end{equation}
Using the above definitions, for the case of $N1$, we get
\begin{equation}
X_{N1}=\left(\begin{array}{c}v_{EW}^2\\0\\0\\0\\0\end{array}\right),\quad
V^\prime_{N1}=\left(\begin{array}{c}0\\m_{\eta^+}^2\\
\frac{1}{2}(\mu_3^2+\lambda_7v_{EW}^2)\\0\\0\end{array}\right)
\end{equation}
While obtaining the form of $V^\prime_{N1}$, we have used the minimization
condition for $N1$,
which is given in Tab. \ref{t2}. Similarly, for other SPs of Eqs. (\ref{n2})$-$(\ref{c11}),
we have obtained $X_{SP}$ and $V^\prime_{SP}$ accordingly. Below we
describe the expression for difference in the value of potential at $N1$ and at any
other SP. To do this computation, we first consider the following products:
$X_{N1}^TV^\prime_{SP}$ and $X_{SP}^TV^\prime_{N1}$. Now, using the quantities
described in the previous paragraph, these products can be expressed as
\begin{eqnarray}
&&X_{N1}^TV^\prime_{SP} = 2V_{N1}-\frac{1}{2}(V_3)_{N1}+X_{N1}^TM_4X_{SP},
\nonumber \\
&&X_{SP}^TV^\prime_{N1} = 2V_{SP}-\frac{1}{2}(V_3)_{SP}+X_{SP}^TM_4X_{N1}
\end{eqnarray}
From the relations in the above equation, we get the following
expression, which gives the relative depth in potential between a SP and $N1$.
\begin{equation}
V_{SP}-V_{N1}=\frac{1}{2}(X_{SP}^TV^\prime_{N1}-X_{N1}^TV^\prime_{SP})
+\frac{1}{4}((V_3)_{SP}-(V_3)_{N1})
\label{difpot}
\end{equation}
It is to remind here that the above expression is obtained after using the
minimization conditions for the SPs of Eqs. (\ref{n1})$-$(\ref{c11}), but
otherwise, these SPs need not be minima.

Using the general expression given in Eq. (\ref{difpot}), we have computed the
differences in the value of potential at $N1$ and at any other SP of
Eqs. (\ref{n2})$-$(\ref{c11}). These expressions are given below.
\begin{eqnarray}
V_{N2}-V_{N1}&=&\frac{v_{\chi(2)}^2}{4}(\mu_3^2+\lambda_7v_{EW}^2),
\\
V_{N3}-V_{N1}&=&\frac{v_{\eta(3)}^2}{2}\mu_2^2-\frac{v_{EW}^2}{2}\mu_1^2,
\label{vn3}
\\
V_{N4}-V_{N1}&=&\frac{v_{\eta(4)}^2}{2}\mu_2^2-\frac{v_{EW}^2}{2}\mu_1^2
+\frac{v_{\chi(4)}^2}{4}\mu_3^2,
\label{vn4}
\\
V_{N5}-V_{N1}&=&\frac{v_{\chi(5)}^2}{4}\mu_3^2-\frac{v_{EW}^2}{2}\mu_1^2,
\label{vn5}
\\
V_{N6}-V_{N1}&=&\frac{v_{\eta(6)}^2}{2}(\mu_2^2+(\lambda_3+\lambda_4+\lambda_5)v_{EW}^2),
\\
V_{N7}-V_{N1}&=&\frac{v_{\eta(7)}^2}{2}m_{\eta_I^0}^2,
\\
V_{N8}-V_{N1}&=&\frac{v_{\eta(8)}^2}{2}(\mu_2^2+(\lambda_3+\lambda_4+\lambda_5)v_{EW}^2)
+\frac{v_{\chi(8)}^2}{4}(\mu_3^2+\lambda_7v_{EW}^2)
\nonumber \\
&&+\frac{Av_{\chi(8)}v_{\eta(8)}}{2v_{\phi(8)}}(v_{EW}^2+v_{\phi(8)}^2),
\label{vn8}
\\
V_{C9}-V_{N1}&=&\frac{c_{\eta(9)}^2+v_{\eta(9)}^2}{2}m_{\eta^+}^2,
\\
V_{C10}-V_{N1}&=&\frac{c_{\eta(10)}^2+v_{\eta(10)}^2}{2}m_{\eta^+}^2,
\\
V_{C11}-V_{N1}&=&\frac{c_{\eta(11)}^2+v_{\eta(11)}^2}{2}m_{\eta^+}^2
+\frac{v_{\chi(11)}^2}{4}(\mu_3^2+\lambda_7v_{EW}^2)
+\frac{1}{2}Av_{\chi(11)}v_{\eta(11)}v_{\phi(11)}
\label{vc11}
\end{eqnarray}
In the above equations, $m_{\eta^+}^2$ and $m_{\eta_I^0}^2$ are mass-square
eigenvalues of $\eta^+$ and $\eta_I^0$ for $N1$, whose expressions are
given in Eq. (\ref{mas-ds}). Now, in the region where $N1$ is a minimum, we should have
$m_{\eta^+}^2>0$ and $m_{\eta_I^0}^2>0$. Hence, in this region, we get
$V_{N7}-V_{N1}>0$, $V_{C9}-V_{N1}>0$ and $V_{C10}-V_{N1}>0$.
This means, if the $N1$ minimum
coexist with either of $N7$, $C9$ and $C10$, the value of potential at $N1$
is always lower than that at $N7$, $C9$ and $C10$. Now, let us consider a
region where the $N1$ minimum coexist with either $N2$ or $N6$. In this region, we should
have $A=0$, since this is one of the minimization conditions for $N2$ and $N6$,
which can be seen from Tab. \ref{t2}. Now, using $A=0$ in Eq. (\ref{mas-ds}) and also
from the fact that $N1$ is a minimum, we get
$\mu_2^2+(\lambda_3+\lambda_4+\lambda_5)v_{EW}^2>0$ and
$\mu_3^2+\lambda_7v_{EW}^2>0$. This implies
$V_{N2}-V_{N1}>0$ and $V_{N6}-V_{N1}>0$, which means that the potential depth at $N1$
is deeper than that at $N2$ and $N6$.

We have argued above that the potential value at $N1$ is always lower than that
at $N2$, $N6$, $N7$, $C9$ and $C10$, in a region where these minima coexist. However,
the situation is different in a region where the $N1$ minimum coexist with either
of $N3$, $N4$ and $N5$. In each of Eqs. (\ref{vn3})$-$(\ref{vn5}), there exist
both positive and negative terms, and hence, it is not guaranteed that the potential
depth at $N1$ is deeper than that at $N3$, $N4$ and $N5$. Now, let us look at
Eqs. (\ref{vn8}) and (\ref{vc11}), which give the relative potential depths for
$N8$ and $C11$ with $N1$. As described in Sec. \ref{s3}, the minima $N8$ and $C11$
should satisfy the conditions of Eq. (\ref{cond}). As a result of this, the
last term of Eqs. (\ref{vn8}) and (\ref{vc11}) should give negative contribution.
Hence, it is not guaranteed that the potential depth at $N1$ is lower than that at
$N8$ and $C11$.

In Sec. \ref{s3}, we have described that eleven different minima can exist in the
Dirac scotogenic model. If we assume that $N1$, which is the desired minimum of this
model, to coexist with other minima of this model, we have shown that the potential
value at $N1$ is not guaranteed to be deeper than that at $N3$, $N4$, $N5$, $N8$ and $C11$.
But otherwise, we have argued that $N1$ minimum is stable against the other vacua of
$N2$, $N6$, $N7$, $C9$ and $C10$. Now, in order to address if $N1$ can be made the
global minimum of this model, we follow the below described steps. First, we
find a parameter region of this model where $N1$ is a minimum. Now in this parameter
region, we check if $N1$ coexist with either of the minima of
$N3$, $N4$, $N5$, $N8$ and $C11$.
If $N1$ minimum is found to coexist with the above described minima, we
demand that the relative potential depths for these minima against $N1$, which are
given in Eqs. (\ref{vn3}), (\ref{vn4}), (\ref{vn5}), (\ref{vn8}) and (\ref{vc11}),
should be positive. After this demand, we see that $N1$ becomes the global
minimum in the region of coexistence. In the next section, we present numerical analysis on
the above described steps and give results on the status of $N1$ as
the global minimum of this model.

\section{Numerical results}
\label{s5}

As we have described at the end of the last section, we first find a parameter region
where $N1$ is a minimum and then check if $N1$ coexist with the other minima of
$N3$, $N4$, $N5$, $N8$ and $C11$ in this region. This parameter region depends on the
dimensionless $\lambda$
parameters and dimensionful parameters of the scalar potential, which is given in
Eq. (\ref{pot}). While finding the above mentioned parameter region, we scan over
these parameters in such a way that perturbativity bounds
on $\lambda$ parameters and boundedness from below conditions on the scalar potential
are satisfied. As a result of this, in our scan over parameters, the below
perturbativity conditions are satisfied on $\lambda$ parameters.
\begin{equation}
|\lambda_i|\leq 4\pi
\label{pc}
\end{equation}
As for the boundedness from below conditions, they are determined by the quartic part
of the scalar potential. In our case, the quartic part of the scalar potential is
same as that considered in \cite{Ferreira:2019iqb}. Hence, the allowed region by
the boundedness from below of the scalar potential of Dirac scotogenic model
is given by \cite{Ferreira:2019iqb}
\begin{eqnarray}
&&\Omega_1\cup\Omega_2,
\nonumber \\
\Omega_1&=&\left\{\lambda_{1,2,6}>0;\sqrt{2\lambda_1\lambda_6}+\lambda_7>0;
\sqrt{2\lambda_2\lambda_6}+\lambda_8>0;\sqrt{\lambda_1\lambda_2}+\lambda_3+D>0;\right.
\nonumber \\
&&\left.\lambda_7+\sqrt{\frac{\lambda_1}{\lambda_2}}\lambda_8\geq 0\right\},
\nonumber \\
\Omega_2&=&\left\{\lambda_{1,2,6}>0;2\lambda_2\lambda_6\geq\lambda_8^2;
\sqrt{2\lambda_1\lambda_6}>-\lambda_7\geq\sqrt{2\lambda_2\lambda_6}+\lambda_8;\right.
\nonumber \\
&&\left.\sqrt{(\lambda_7^2-2\lambda_1\lambda_6)(\lambda_8^2-2\lambda_2\lambda_6)}>
\lambda_7\lambda_8-2(D+\lambda_3)\lambda_6\right\},
\nonumber \\
D&=&{\rm min}(\lambda_4-|\lambda_5|,0)
\label{bfbc}
\end{eqnarray}

In our numerical analysis, we randomly generate the $\lambda$ parameters
in such a way that the bounds in Eqs. (\ref{pc}) and (\ref{bfbc}) are satisfied.
In addition to these
$\lambda$ parameters, dimensionful parameters are also exist in the scalar
potential of Eq. (\ref{pot}). Among these dimensionful parameters, we fix
$\mu_1^2=-\lambda_1v_{EW}^2$ in our numerical analysis. This is due to the fact that
we search for a region where $N1$ is a minimum and the above relation is the
minimization condition for this minimum. Rest of the dimensionful parameters
such as $\mu_2^2,\mu_3^2,A$ are either fixed to a value or randomly generated.
Here, we notice that the $A$-parameter is necessary in order to
explain neutrino masses in the Dirac scotogenic model, which we have discussed
in Sec. \ref{s2}. It is described that this parameter breaks $Z_2^{(A)}$ symmetry
softly, and hence, this parameter should be small. As a result of this, we have
taken this parameter to be small in our numerical analysis. On the other hand,
the parameters $\mu_{2,3}^2$, except that for $N1$, determine the minimization
conditions of all minima, which can be seen from Tab. \ref{t2}. As a result of
this, either we have fixed these parameters in order to satisfy minimization
conditions of a particular minimum, or else, we have generated them randomly.
We discuss details on these parameters later.

In the previous paragraph, we have described our methodology in scanning over
parameters of the scalar potential. We notice that these
parameters are physical, since they can be measured in future experiments.
Apart from these parameters, in our scanning procedure, we have randomly
varied the non-zero VEVs of Eqs. (\ref{n2})$-$(\ref{c11}). It is to remind
here that the minima of Eqs. (\ref{n2})$-$(\ref{c11}) are unphysical, since
they do not correspond to the vacuum of our physical world. As a result of
this, the non-zero VEVs of Eqs. (\ref{n2})$-$(\ref{c11}) are unknown, and
a priori, they can take arbitrary values.

As described previously, we scan over parameters of the model and find a region
where $N1$ is a minimum. For the $N1$ minimum,
the mass eigenstates of the scalar fields are physical, since this minimum
corresponds to the vacuum of our physical world. The mass eigenstates of this
minimum are described in Eq. (\ref{mas-ds}).
Among these eigenstates, only the Higgs boson is found in the LHC experiment
and rest of the scalar fields are yet to be found. In order to fit the Higgs boson
mass in our analysis, we have taken $m_H=$ 125.25 GeV \cite{Workman:2022ynf}.
This value of Higgs boson mass also fixes the $\lambda_1$ parameter in our analysis.
Regarding the additional mass eigenstates of the $N1$ minimum, they should satisfy
the phenomenological lower bounds on their masses
due to non-observation of these fields in collider experiments. As a result of
this, we apply the following bounds on these masses in our analysis.
\begin{equation}
m_{\eta^+}>105~{\rm GeV},\quad m_{\zeta_{1,2,3}}>5~{\rm GeV}
\end{equation}
The lower bound of 105 GeV on $m_{\eta^+}$ is based on the fact that no non-SM
charged particle is found in the LEP collider.

Below we present our numerical results in the form of percentage of chances on the
coexistence of $N1$ minimum with other minima. To get these results, we have done
multiple scans over the parameter space. In each of these scans, the percentages
are computed
after generating a large number of parametric points (at least 15000), which satisfy
the bounds of Eqs. (\ref{pc}) and (\ref{bfbc}).

In Tab. \ref{t3}, we have given numerical results on coexistence of the
$N1$ and $N3$ minima.
\begin{table}[!h]
\centering
\begin{tabular}{|c|c|c|c|} \hline
$\frac{A}{v_{EW}}$ & $N1$ & $N1$ and $N3$ & $N1$ deeper than $N3$ \\ \hline
0.1 & 43.3$\%$ & 15.2$\%$ & 11.1$\%$ \\
$10^{-3}$ & 43.6$\%$ & 14.7$\%$ & 10.6$\%$ \\
$10^{-5}$ & 43.7$\%$ & 15.1$\%$ & 11.0$\%$ \\ \hline
\end{tabular}
\caption{For various values of $A$, percentages of coexistence between the minima
$N1$ and $N3$ are given. In the second column, percentage of existence of
$N1$ minimum is given. In the third column, percentage of coexistence of $N1$ and
$N3$ is given. In the fourth column, percentage of coexistence of $N1$ and
$N3$ with $V_{N1}<V_{N3}$ is given.}
\label{t3}
\end{table}
In order to
search for this coexistence, in our scan over parameters of the scalar potential,
we satisfy the minimization conditions only for $N1$ and $N3$. As a result of this,
from Tab. \ref{t2}, we notice that the relations for $\mu_1^2$ and $\mu_2^2$ are fixed,
but $\mu_3^2$ can be chosen arbitrarily. Moreover, the quantity $v_{\eta(3)}$ can
also be chosen arbitrarily. In our scan, we have varied $v_{\eta(3)}$ and $\mu_3^2$
randomly in the ranges of $(-100,100)$ GeV and
$(-10^{5},10^5)$ GeV$^2$, respectively. For these ranges of
$v_{\eta(3)}$ and $\mu_3^2$, we have found that the percentages given in
Tab. \ref{t3} to be large. In Tab. \ref{t3}, we have chosen $A$-parameter to be
suppressed as compared to the electroweak scale, since this parameter should be
small, which is described previously. From this table, we see that nearly 44$\%$ chance
is there in finding the $N1$ minimum. On the other hand, around 15$\%$ chance is there in
finding a parametric point for which both $N1$ and $N3$ are minima. From the last
column of Tab. \ref{t3}, we see that
around 11$\%$ of points correspond to the $N1$ minimum to be deeper than the
$N3$ minimum in terms of potential depth. In this table, the difference of percentages
between the second and third columns, for a particular value of $A$, is about 29$\%$.
This percentage corresponds to the parametric region in which $N1$ is the
only minimum. We see
here that, out of the total area of scan, in a significant fraction of it, $N1$
is the only minimum. From this table, we see that the numerical
results are not sensitive to the $A$-parameter. Moreover, by changing the
sign of $A$, we nearly got the same percentages that are given in Tab. \ref{t3}.

After satisfying only the minimization conditions for $N1$ and $N4$ in our
scanning procedure, we have
obtained numerical results on the coexistence of these minima. These results are given
in Tab. \ref{t4}.
\begin{table}[!h]
\centering
\begin{tabular}{|c|c|c|c|} \hline
$\frac{A}{v_{EW}}$ & $N1$ & $N1$ and $N4$ & $N1$ deeper than $N4$ \\ \hline
0 & 33.5$\%$ & 18.3$\%$ & 12.3$\%$ \\ \hline
\end{tabular}
\caption{Percentages of coexistence between the minima $N1$ and $N4$ are given.
The columns in this table are analogous to that in Tab. \ref{t3}.}
\label{t4}
\end{table}
In a region where $N1$ and $N4$ coexist, the parameters $\mu_{2,3}^2$ are
fixed according to the relations given in Tab. \ref{t2}. Here, the quantities
$v_{\eta(4)}$ and $v_{\chi(4)}$ can be chosen arbitrarily. We have varied these
quantities independently in the range $(-100,100)$ GeV in order to obtain percentages in
Tab. \ref{t4}. For this particular range of values, the percentages given
in this table are large. We notice that the minima $N1$ and $N4$ coexist in a
region where $A=0$. For this value of $A$, and from the discussion given below
Eq. (\ref{mnu}), the neutrino masses in this model become zero. Hence, in order
to explain non-zero masses to neutrinos, we should choose $A\neq0$, and thereby,
the coexistence between $N1$ and $N4$ can be avoided.

In analogy to the results described for Tabs. \ref{t3} and \ref{t4}, we have
searched for the coexistence of the minima $N1$ and $N5$. The results of this
coexistence are given in Tab. \ref{t5}.
\begin{table}[!h]
\centering
\begin{tabular}{|c|c|c|c|} \hline
$\frac{A}{v_{EW}}$ & $N1$ & $N1$ and $N5$ & $N1$ deeper than $N5$ \\ \hline
0.1 & 36.0$\%$ & 9.4$\%$ & 7.8$\%$ \\
$10^{-3}$ & 36.0$\%$ & 9.5$\%$ & 8.1$\%$ \\
$10^{-5}$ & 36.4$\%$ & 9.3$\%$ & 8.0$\%$ \\ \hline
\end{tabular}
\caption{Percentages of coexistence between the minima $N1$ and $N5$ are given.
The columns in this table are analogous to that in Tab. \ref{t3}.}
\label{t5}
\end{table}
In order to get results in this table, we have varied
$v_{\chi(5)}$ and $\mu_2^2$ in the ranges
$(-100,100)$ GeV and $(-10^{5},10^5)$ GeV$^2$, respectively. From this
table, we see that the percentages are not sensitive to the value of $A$.
Moreover, we have seen that these results are not sensitive to the sign of $A$.
After comparing the
results in Tabs. \ref{t3} and \ref{t5}, we see that the percentage of
coexistence between $N1$ and $N5$ is lower as compared to that between
$N1$ and $N3$.

In Tab. \ref{t6}, we have given the results on the coexistence of the minima
$N1$ and $N8$.
\begin{table}[!h]
\centering
\begin{tabular}{|c|c|c|c|} \hline
$\frac{|A|}{v_{EW}}$ & $N1$ & $N1$ and $N8$ & $N1$ deeper than $N8$ \\ \hline
0.1 & 26.4$\%$ & 4.8$\%$ & 4.2$\%$ \\
$10^{-3}$ & 21.0$\%$ & 0.57$\%$ & 0.56$\%$ \\
$10^{-5}$ & 21.4$\%$ & 0.14$\%$ & 0.14$\%$ \\
0 & 21.7$\%$ & 0$\%$ & 0$\%$ \\ \hline
\end{tabular}
\caption{Percentages of coexistence between the minima $N1$ and $N8$ are given.
The columns in this table are analogous to that in Tab. \ref{t3}.}
\label{t6}
\end{table}
These results are obtained after satisfying only the minimization conditions of
$N1$ and $N8$. Among these conditions, the relation for $\mu_1^2$ gives a
constraint relation. This relation is satisfied by solving for unknown
$\lambda$ parameters and later we have checked if these $\lambda$ parameters obey
the conditions of Eqs. (\ref{pc}) and (\ref{bfbc}).
While obtaining the results in Tab. \ref{t6}, we have
varied each of $v_{\phi(8)},v_{\eta(8)},v_{\chi(8)}$ independently in the range
$(-100,100)$ GeV, for which the percentages in this table are found to be large.
After comparing the results in this table with that of Tabs. \ref{t3} and \ref{t5},
we see that the percentage of coexistence for $N1$ and $N8$ is lower than that
for $N1$ with either $N3$ or $N5$.
In Tab. \ref{t6}, we have given the value of $|A|$, since the sign of $A$ is determined
by the sign of $v_{\phi(8)}v_{\eta(8)}v_{\chi(8)}$ in our scanning procedure. This
is due to the fact that we have a condition of Eq. (\ref{cond}) for the minimum $N8$.
We notice that by decreasing the value of $|A|$, the percentage for coexistence
between the minima $N1$ and $N8$ is getting decreased. We have found that,
for $A=0$, the minimum $N8$ exist in some region of parameter space. However,
$N8$ minimum does not coexist with $N1$ minimum in the region where $A=0$.
In the numerical analysis, we have noticed that $N8$ becomes a saddle point
in the region where $N1$ is a minimum, for $A=0$.

In our scanning procedure, we have searched
for the coexistence of minima $N1$ and $C11$. While satisfying the minimization
conditions of these minima, the relation for $\mu_1^2$ gives a constraint relation.
This is solved in an analogous way of what we have described for Tab. \ref{t6}.
We have varied the quantities $v_{\phi(11)},c_{\eta(11)},v_{\eta(11)},v_{\chi(11)}$
arbitrarily in our analysis. Now, we see that the parameter $A$ is determined
by the above quantities through one of the minimization conditions of $C11$,
which is given in Tab. \ref{t2}. As already described before, the parameter $A$
should be small in order to be consistent with the model framework. As a result
of this, we have demanded $\frac{|A|}{v_{EW}}\leq0.1$ in our analysis, and thereafter,
we have not found a region for the coexistence of the minima $N1$ and $C11$.
In the appendix \ref{app}, we argue that for $A$ to be a negligibly small
variable, $C11$ becomes a saddle point in the region where $N1$ is a minimum.
The result shown in this appendix, concurs with our numerical result that
the minima $N1$ and $C11$ do not coexist for $A$ to be a small variable.
On the other hand, in our analysis, for $\frac{|A|}{v_{EW}}>0.1$,
we have found a region for the coexistence of minima
$N1$ and $C11$. However, this region is not viable due to the above mentioned
reasons.

In the previous section we have argued that, even if the minimum $N1$ coexist
with either of the minima $N2$, $N6$, $N7$, $C9$ or $C10$, the potential depth
at $N1$ is always deeper
than that at the other minima mentioned here. For the sake of completeness,
in our numerical analysis, we have searched if $N1$ coexist with any of the
above mentioned minima. This searched is done in an analogous way of what we have
described for the results of Tabs. \ref{t3}$-$\ref{t6}. In our analysis, we
have not found a region where $N1$ coexist with either of the minima
$N2$, $N6$, $N7$, $C9$ or $C10$. In the appendix \ref{app}, we argue that
some of these SPs become saddle points in a region where $N1$ is a minimum.

From the numerical results presented so far, we have seen that in the Dirac
scotogenic model, the minimum $N1$ can coexist with either of the minima
$N3$, $N4$, $N5$ or $N8$. Here, it should be noted that the coexistence of
the minima $N1$ and $N4$ happen in a region where $A=0$. This region is not
interesting, since neutrino masses become zero in this region of the model.
On the other hand, the coexistence of $N1$ with the other minima of
$N3$, $N5$ and $N8$ can happen in a region where $A\neq0$, which is an interesting
region to us from the point of neutrino masses. Hence, in this region, we have
searched to see if $N1$ minimum coexist with more than one minima of $N3$, $N5$ and $N8$.
In Tab. \ref{t7}, we have given results on the coexistence of the minima among
$N1$, $N3$ and $N5$.
\begin{table}[!h]
\centering
\begin{tabular}{|c|c|c|c|} \hline
$\frac{A}{v_{EW}}$ & $N1$ & $N1$, $N3$ and $N5$ & $N1$ deeper than $N3$, $N5$ \\ \hline
0.1 & 32.9$\%$ & 0.81$\%$ & 0.77$\%$ \\ \hline
\end{tabular}
\caption{Percentages of coexistence among the minima $N1$, $N3$ and $N5$ are given.
The columns in this table are analogous to that in Tab. \ref{t3}.}
\label{t7}
\end{table}
While obtaining results on this coexistence, we have satisfied minimization
conditions only for these minima. Also, for these results, we have varied
$v_{\eta(3)}$ and $v_{\chi(5)}$ independently in the range $(-100,100)$ GeV.
We see that the percentage of coexistence among these minima is far less than
that given in Tabs. \ref{t3} and \ref{t5}. We have noticed that the percentages
given in Tab. \ref{t7} are not sensitive to the value of $A$, which is also the
case in Tabs. \ref{t3} and \ref{t5}.

We have also searched for the coexistence of minima among $N1$, $N3$ and $N8$
and also among $N1$, $N5$ and $N8$. In our analysis, we have found that
the above mentioned coexistences can happen in a region of $\frac{|A|}{v_{EW}}>0.1$.
However, this region is not viable since $A$ should be a small parameter. Hence, we have
have demanded $\frac{|A|}{v_{EW}}\leq0.1$ in our analysis, and thereafter, we have
not found a region for the above mentioned coexistences. This result may be
understood in the following way. In Tab. \ref{t6}, it is shown that in the
limit that $A$ is small, the percentage of chances for $N8$ to coexist with $N1$
is getting decreased. Hence, for small $A$, it is difficult for $N8$ to coexist
with $N1$ and $N3$ or with $N1$ and $N5$. Finally, we have
found that the minima $N1$, $N3$, $N5$ and $N8$ can coexist in a region of
$\frac{|A|}{v_{EW}}>0.1$, but otherwise, we have not found a region for these
minima to coexist.

While describing the results of Tab. \ref{t3}, we have mentioned that the
difference in the percentages of second and third columns corresponds to the
fraction of total scanned region, where $N1$ is the only minimum. This statement
is true even for Tabs. \ref{t4}$-$\ref{t7}. In any of these tables, the difference
in the percentages of second and third columns, for a particular value of $A$,
is at least 15$\%$. After comparing percentages in these tables, we notice that
in a significant parameter region of our scanning process, $N1$ is the only minimum.
Moreover, in this region, $N1$ is obviously the global minimum of the Dirac
scotogenic model. On
the other hand, there exist a certain parameter region where $N1$ can coexist
with other minima, whose percentage of chances of finding is given by the third column of
Tabs. \ref{t3}$-$\ref{t7}. In the region where $N1$ coexist with other minima,
we have demanded that the potential depth at $N1$ is lower than that at the other
minima, so that $N1$ can be the global minimum of this model. The fourth column
of Tabs. \ref{t3}$-$\ref{t7} gives the percentage of chances for $N1$ to be the
global minimum in the region where it coexist with other minima.

\section{Higgs to diphoton decay}
\label{s6}

In the previous section, we have described the status of the $N1$ minimum as the
global minimum of the Dirac scotogenic model. We have noticed that the analysis
of the previous section is determined by the parameters of the scalar potential.
We now want to study the impact of this analysis on phenomenological observable
quantities. One of the observable quantities is the decay $H\to\gamma\gamma$ upon
which the scalar sector of this model can have an impact. This decay is in general driven
by charged particles through a loop induced process. In the Dirac scotogenic
model, this decay gets additional contribution due to $\eta^+$ field. This decay
is the subject of experimental investigation, since it can distinguish any new
physics signal from that of SM. As part of this investigation, in the
LHC experiment, the signal strength of $H\to\gamma\gamma$ is measured and it is found to be
$1.1\pm0.07$ \cite{Workman:2022ynf}. In this section, we compute the signal
strength of this decay in the Dirac scotogenic model and study consequences
on this quantity due to analysis of the previous section.

The signal strength of $H\to\gamma\gamma$ is defined
as the ratio of the observed cross section of $pp\to H\to\gamma\gamma$ against to
the same quantity computed in the SM. The observed cross section of
$pp\to H\to\gamma\gamma$ in the LHC experiment should match with that computed
in the model of our work, which is the Dirac scotogenic model. We notice
that the production cross section for the Higgs boson in the Dirac scotogenic
model is nearly same as that in the SM, since the dominant process for this production
is through the gluon fusion. As a result of this, after using the narrow width
approximation, the signal strength of $H\to\gamma\gamma$ in our work is given by
\begin{equation}
R_{\gamma\gamma}=\frac{Br(H\to\gamma\gamma)_{DSM}}{Br(H\to\gamma\gamma)_{SM}}=
\frac{\Gamma(H\to\gamma\gamma)_{DSM}}{\Gamma(H\to\gamma\gamma)_{SM}}
\frac{\Gamma_{SM}^H}{\Gamma_{DSM}^H}
\label{rgg}
\end{equation}
Here, the quantities having the suffixes DSM and SM are the ones computed
in the Dirac scotogenic and standard models, respectively. $\Gamma^H_{DSM,SM}$
correspond to the total decay widths of the Higgs boson in the above two
models. The decay widths of $H\to\gamma\gamma$, which are required in
Eq. (\ref{rgg}), are computed using a general expression for this quantity
given in \cite{Shifman:1979eb}. For the case of Dirac scotogenic model, we have
\begin{eqnarray}
\Gamma(H\to\gamma\gamma)_{DSM}&=&\frac{\alpha^2G_Fm_H^3}{128\sqrt{2}\pi^3}
\left|\sum_f N_fQ_f^2F_{1/2}(\beta_f)+
F_1(\beta_W)
+\frac{\lambda_3v_{EW}^2}{m^2_{\eta^+}}
F_0(\beta_\eta)\right|^2,
\nonumber \\
&&\quad \beta_f=\frac{4m_f^2}{m_H^2},\quad \beta_W=\frac{4m_W^2}{m_H^2},\quad
\beta_\eta=\frac{4m_{\eta^+}^2}{m_H^2},
\label{dwhgg}
\end{eqnarray}
where $\alpha$ and $G_F$ are fine-structure and Fermi constants, respectively.
Here, $N_f$, $Q_f$ and $m_f$ are color factor, charge and mass of SM fermion,
respectively. After excluding the last term in the modulus of the above equation,
we get the expression for $\Gamma(H\to\gamma\gamma)_{SM}$. The $F$-functions
in Eq. (\ref{dwhgg}) are the form factors of spin-1/2, -1 and -0 fields, which
drive the decay $H\to\gamma\gamma$. These functions are given below.
\begin{eqnarray}
F_{1/2}(\beta)&=&-2\beta[1+(1-\beta)f(\beta)],
\nonumber \\
F_1(\beta)&=&2+3\beta+3\beta(2-\beta)f(\beta),
\nonumber \\
F_0(\beta)&=&\beta[1-\beta f(\beta)],
\nonumber \\
f(\beta)&=&\left\{\begin{array}{cr}
\left(\sin^{-1}\frac{1}{\sqrt{\beta}}\right)^2, & \beta\geq1 \\
-\frac{1}{4}\left[\ln\frac{1+\sqrt{1-\beta}}{1-\sqrt{1-\beta}}-i\pi\right]^2, & \beta<1
\end{array}\right.
\end{eqnarray}

While computing $R_{\gamma\gamma}$ in our analysis, we have taken the total
decay width of Higgs boson in the SM as $\Gamma_{SM}^H=4.1\times10^{-3}$
GeV \cite{LHCHiggsCrossSectionWorkingGroup:2013rie}. Now, the total decay
width of Higgs boson in the Dirac scotogenic model, to a leading order, is given by
\begin{equation}
\Gamma_{DSM}^H=\Gamma_{SM}^H+\Gamma(H\to\zeta_1\zeta_1)+\Gamma(H\to\zeta_1\zeta_2)
+\Gamma(H\to\zeta_2\zeta_2)+\Gamma(H\to\zeta_3\zeta_3)+\Gamma(H\to\eta^+\eta^-)
\end{equation}
In the above equation, the partial decay widths of Higgs boson into scalar
particles of Dirac scotogenic model are computed using tree level couplings
of these processes. These couplings are given below.
\begin{eqnarray}
&&C_{H\zeta_1\zeta_1}=
-i\sqrt{2}v_{EW}[(\lambda_3+\lambda_4+\lambda_5)O_{12}^2+\lambda_7O_{22}^2]-i2AO_{12}O_{22},
\nonumber \\
&&C_{H\zeta_1\zeta_2}=
-i\sqrt{2}v_{EW}[(\lambda_3+\lambda_4+\lambda_5)O_{11}O_{12}+\lambda_7O_{21}O_{22}]
-iA(O_{12}O_{21}+O_{11}O_{22}),
\nonumber \\
&&C_{H\zeta_2\zeta_2}=
-i\sqrt{2}v_{EW}[(\lambda_3+\lambda_4+\lambda_5)O_{11}^2+\lambda_7O_{21}^2]-i2AO_{11}O_{21}
\nonumber \\
&&C_{H\zeta_3\zeta_3}=
-i\sqrt{2}(\lambda_3+\lambda_4-\lambda_5)v_{EW}
\nonumber \\
&&C_{H\eta^+\eta^-}=-i\sqrt{2}\lambda_3v_{EW}
\label{coup}
\end{eqnarray}
Here, $O_{ij}$, where $i,j=1,2$, is the element of an orthogonal matrix which diagonalizes
the mixing mass matrix of $\eta_R^0$ and $\chi$, which is given in Eq. (\ref{mas-ds}).
In our numerical analysis, we follow the below convention for the
diagonalization of this mixing matrix.
\begin{equation}
O^T M_{\eta_R^0\chi}^2O={\rm diag}(m_{\zeta_2},m_{\zeta_1})
\end{equation}
Here we identify $m_{\zeta_1}$ to be the lightest.

\begin{figure}[!h]
\centering

\includegraphics[width=3.0in]{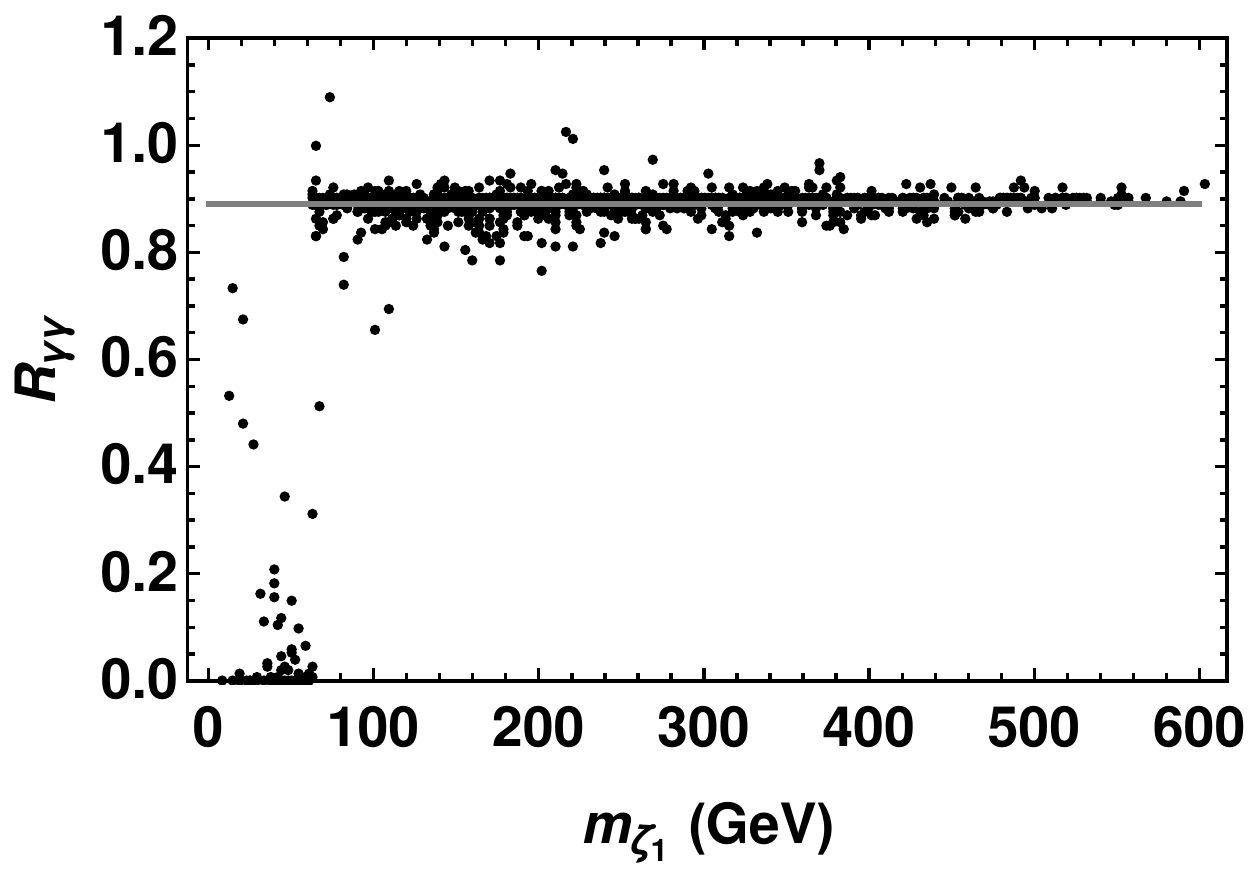}
\includegraphics[width=3.0in]{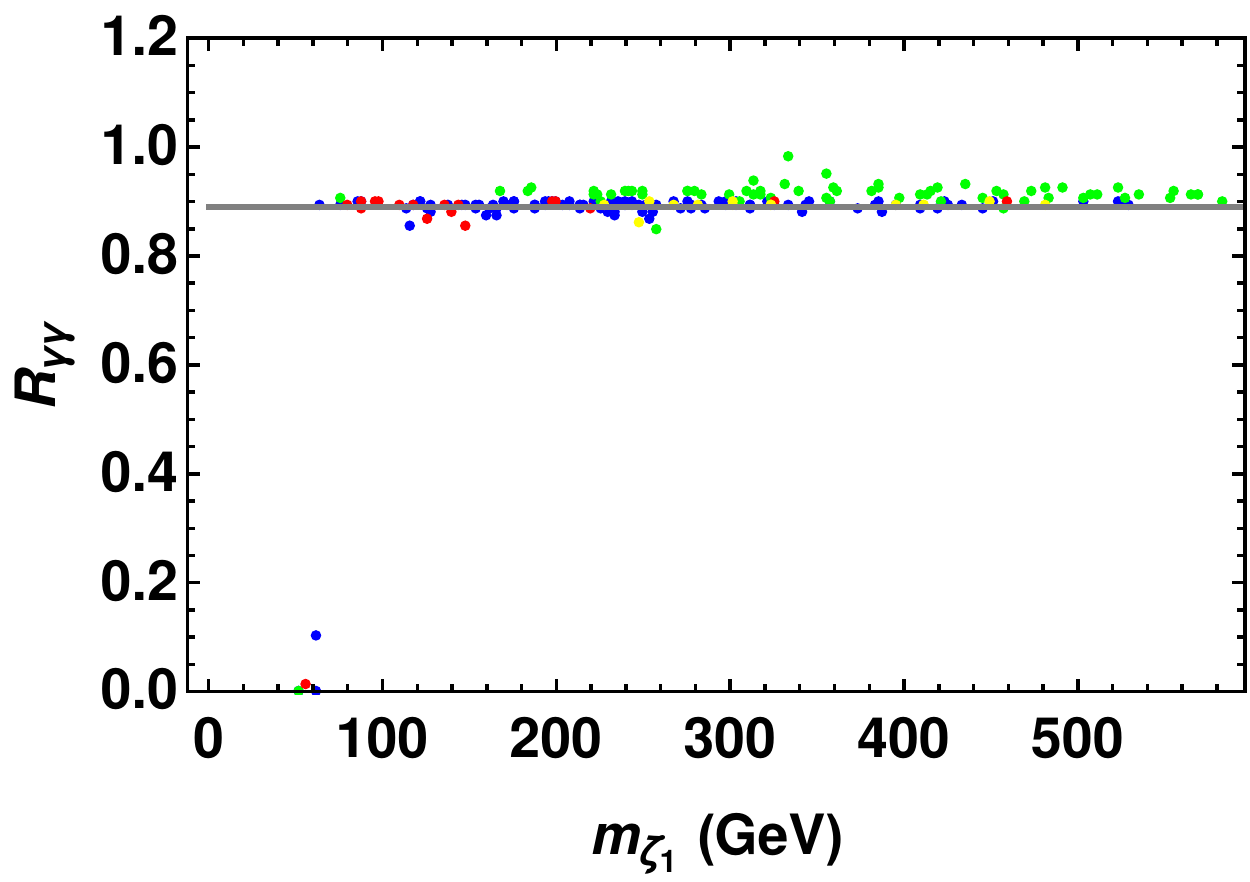}

\caption{The left-hand side plot is for parametric points for which $N1$ is the
only minimum. Right-hand side plot is for parametric points where the $N1$
minimum coexist with other minima of the model, and moreover, for these points, the potential
depth at $N1$ is deeper than that at other minima. In the right-hand side
plot, blue, green, red and yellow are the points where $N1$ minimum coexist
with $N3$, $N5$, $N8$ and $N3+N5$ minima respectively. The horizontal line in
these plots indicates the experimentally allowed lower 3$\sigma$ value of
$R_{\gamma\gamma}$. In these plots, we have taken $\frac{|A|}{v_{EW}}$ = 0.01.}
\label{f1}
\end{figure}

Using the expressions, which are described above, we have computed $R_{\gamma\gamma}$
after scanning over parameters of the scalar potential of Dirac scotogenic model.
We notice here that while scanning over these parameters, it is possible that the
$N1$ minimum of this model may coexist with other minima of the model. In order
to see the effect of this coexistence, we have done the scanning in a way of what we
have described in the previous section. The difference in the scanning of previous
section and the current section is that,
in the current analysis, we have taken $m_{\zeta_1}$ as the lightest among the masses of
additional particles of Dirac scotogenic model. It is to remind here
that the additional particles of this model are charged under $Z_2^{(B)}$
symmetry, which is exact. Hence, $\zeta_1$ is a possible candidate for the
dark matter. As a result of this, we have applied the following constraints in
the current analysis.
\begin{equation}
m_{\eta^+}>105~{\rm GeV},\quad 5~{\rm GeV}<m_{\zeta_1}<m_{\zeta_{2,3}},m_{\eta^+}
\end{equation}
The results of our analysis, after doing a generic scan over parameters,
are given in Fig. \ref{f1}.

We see that many points exist in the left-hand side plot as compared to that in
the right-hand side plot of Fig. \ref{f1}. This implies that many parametric points
in our scan correspond to the points where $N1$ is the only minimum. This result
is also described in the numerical analysis of previous section. From the plots
of Fig. \ref{f1}, we see that for $m_{\zeta_1}<m_H/2$, $R_{\gamma\gamma}$ is
suppressed. This suppression is due to the factor $\frac{\Gamma_{SM}^H}{\Gamma_{DSM}^H}$
in $R_{\gamma\gamma}$. For $m_{\zeta_1}<m_H/2$, the decay channel $H\to\zeta_1\zeta_1$
opens up, whose decay width is found to be at least about 0.1 GeV, which gives
the necessary suppression in the above mentioned factor. Also, in Fig. \ref{f1},
we have found that the points, for which $m_{\zeta_1}<m_H/2$, do not satisfy
the constraint due to invisible decay of Higgs boson. For $m_{\zeta_1}<m_H/2$,
the Higgs boson of this model decays invisibly, whose branching ratio is
constrained to be $Br_{H\to inv}<0.145$ \cite{ATLAS:2022yvh}.
In order to get
enhancement of $R_{\gamma\gamma}$ for $m_{\zeta_1}<m_H/2$, one has to suppress
the couplings of Higgs to scalar particles, whose expressions are given in
Eq. (\ref{coup}). In the scanning process, after including the above mentioned
constraint on $Br_{H\to inv}$, we have seen that the
suppression in couplings is possible, and thereby,
$R_{\gamma\gamma}$ can be enhanced to within the experimentally allowed region,
for $m_{\zeta_1}<m_H/2$. On the other hand, for $m_{\zeta_1}>m_H/2$, there
won't be suppression in $\frac{\Gamma_{SM}^H}{\Gamma_{DSM}^H}$. Hence, majority
of the points are within the experimentally allowed region, for $m_{\zeta_1}>m_H/2$.

In both the plots of Fig. \ref{f1}, for $m_{\zeta_1}>m_H/2$, the points are
around the horizontal line, which corresponds to lower 3$\sigma$ allowed
value of $R_{\gamma\gamma}$. Moreover, we see that only few points give the
enhancement of $R_{\gamma\gamma}>1$. In this regard, see \cite{Arhrib:2012ia,
Swiezewska:2012eh,Akeroyd:2012ms,Hundi:2013lca,Yue:2013qba,Garcia-Jimenez:2017ezn}
where enhancement of $R_{\gamma\gamma}>1$ has been reported in various models.
Below we describe the reasons for not getting much enhancement in
$R_{\gamma\gamma}$ in the plots of Fig. \ref{f1}. As already explained before,
for $m_{\zeta_1}>m_H/2$, we get $\frac{\Gamma_{SM}^H}{\Gamma_{DSM}^H}=1$. Hence,
in the region of $m_{\zeta_1}>m_H/2$, we get $R_{\gamma\gamma}>1$ only if
$\lambda_3<0$, since $\beta_\eta>1$. The value of $\lambda_3$ is constrained
by the bounded from below conditions of Eq. (\ref{bfbc}).
After satisfying these conditions, we have noticed that $\lambda_3>-2$. Since
the negative values of $\lambda_3$ are restricted by these conditions, we do
not get much enhancement in $R_{\gamma\gamma}$ in our work. On the other hand,
in our scanning procedure, we have seen that
by restricting $m_{\eta^+}$ to as low as 130 GeV,
it is possible for $R_{\gamma\gamma}$ to be as high as 1.3.

In the region where $N1$ minimum coexist with $N3$ minimum, we do not get
$R_{\gamma\gamma}>1$. This is due to the fact that, in this region we have
$\mu_2^2<0$, because of the minimization condition of $N3$. As a result of this,
in order to get $m^2_{\eta^+}>0$ we should have $\lambda_3>0$, and thereby we
get $R_{\gamma\gamma}<1$. Similarly, in the region where $N1$ and $N8$ minima coexist,
we have found $\lambda_3>0$ in the scanning analysis, and hence, we get
$R_{\gamma\gamma}<1$. On the other hand, in the region where $N1$ and $N5$
minima coexist, $\mu_2^2$ is a free parameter. Hence, in this region, we
can choose $\mu_2^2>0$, and thereby $\lambda_3$ can be negative, so that
$R_{\gamma\gamma}>1$. Now, it should be clear that in a region where $N1$ is
the only minimum, we can get $\mu_2^2>0$, and thereby, we get $R_{\gamma\gamma}>1$.
From the above described results, we see that the future determination of
$R_{\gamma\gamma}$ by the LHC experiment can have an impact on the study of
global minimum of this model. It is to be noted that a precise determination of
$R_{\gamma\gamma}$ in the LHC experiment can distinguish the above described
vacuum realizations. On the other hand, if the error bar is large enough so that the
allowed value of $R_{\gamma\gamma}$ is around 1, the above vacuum realizations
may not be distinguished.
Finally, from Fig. \ref{f1}, we notice that the experimentally allowed values of
$R_{\gamma\gamma}$ can be fitted in the Dirac scotogenic model, irrespective
of the fact that the $N1$ minimum either coexist or not with other minima of this model.

\section{Possibility of a scalar dark matter}
\label{s7}

It is described in Sec. \ref{s2} that, due to exact symmetry of $Z_2^{(B)}$,
the lightest among the additional particles of the Dirac scotogenic model
is a candidate for dark matter. We notice that the lightest among the singlet
Dirac fermions $N^D_k$ is a possible candidate for dark matter. In the
original model \cite{Farzan:2012sa}, this possibility has been studied and it
is shown that $N^D_1$ can consistently explain the dark matter phenomenology.
See \cite{Ghosh:2022fws}, for another work in this direction.
In the present work, since our motivation
is to study the scalar sector of the Dirac scotogenic model, we study
on the possibility of scalar dark matter in this model.

In the previous section, it is described that,
while obtaining results of Fig. \ref{f1}, we have taken $\zeta_1$ as the
lightest particle among the additional particles of Dirac scotogenic
model. From Eq. (\ref{mas-ds}), we notice that $\zeta_1$ is an admixture
of $\eta_R^0$ and $\chi$. For sufficiently small $A$, which is the case in
Fig. \ref{f1}, $\eta_R^0$ and $\chi$
are nearly equal to the mass eigenstates of this model. In such a case,
$\zeta_1$ is dominantly made of either $\eta_R^0$ or $\chi$, depending on
the parameter choice of the model. We see that the scalar dark matter in this
model is either part of an $SU(2)_L$ doublet or a singlet field. We first consider the
case of $\zeta_1$ being dominantly made of $\chi$ and analyze if this case
can consistently explain all the dark matter phenomenology. Later we
analyze the case where $\zeta_1$ is dominantly made of $\eta_R^0$.

In the phenomenology of dark matter, we need to explain the relic abundance
of dark matter in the present universe and also the null results of direct and
indirect searches for dark matter detection.
The current relic density of dark matter is 0.12$\pm$0.0012 \cite{Planck:2018vyg}.
In order to explain this relic density, we need to compute thermally averaged
pair-annihilation cross section
of dark matter times the relative velocity of these particles, which is
denoted by $\langle\sigma v_{rel}\rangle$. To a good approximation, for a cold
dark matter, the above
mentioned relic density can be fitted if $\langle\sigma v_{rel}\rangle\approx3\times
10^{-26}$ cm$^3$/s = 1 pb \cite{Workman:2022ynf}. In the Dirac scotogenic model,
for the case of dark matter $\zeta_1\approx\chi$, the possible pair-annihilation
processes at tree level are as follows:
\begin{eqnarray}
&&\zeta_1\zeta_1\to\nu^D_i\bar{\nu}^D_j,
\label{t-chan} \\
&&\zeta_1\zeta_1\to HH,
\label{quat} \\
&&\zeta_1\zeta_1\to H^*\to f\bar{f},W^+W^-,ZZ,HH
\label{tri}
\end{eqnarray}
Here, $H^*$ is virtual Higgs boson and $f$ is any SM fermion. $\nu^D_i$ are
the mass eigenstates of the neutrino fields of the Dirac scotogenic model.

The reaction in Eq. (\ref{t-chan}) is driven by
the $h_{k\alpha}$ couplings of Eq. (\ref{lag}), via a $t$-channel process mediated
by the singlet fermions $N^D_k$. We can take the mass of $N^D_k$ ($m_{N_k}$) to be around
1 TeV and the couplings $h_{k\alpha}\sim1$. For these values of $m_{N_k}$ and
$h_{k\alpha}$, and from the discussion given below Eq. (\ref{mnu}), we see that
it is possible to explain small masses to neutrinos either
by suppressing the $A$-parameter or by taking degenerate masses to $\zeta_{1,2}$.
Now the annihilation cross section for the process in Eq. (\ref{t-chan}) is given by
\begin{equation}
\sigma(\zeta_1\zeta_1\to\nu^D_i\bar{\nu}^D_j)v_{rel}=
\frac{1}{8\pi}\sum_k
\frac{|\sum_{\alpha,\beta}h_{k\alpha}V^*_{\alpha i}h^*_{k\beta}V_{\beta j}|^2}
{(m_{\zeta_1}^2+m_{N_k}^2)^2}
(m_{\nu_i}^2+m_{\nu_j}^2)
\label{paircs}
\end{equation}
Here, $V_{\alpha i}$ are the elements of $V$, which diagonalize the neutrino
mass matix, which is discussed in Sec. \ref{s2}. For neutrino
masses around 0.1 eV, the above pair-annihilation cross section is
suppressed by fourteen orders of magnitude as compared to the required amount. Hence,
the process in Eq. (\ref{t-chan}) cannot explain the relic density of dark matter.

The processes in Eqs. (\ref{quat}) and (\ref{tri}) are driven due to quartic and
trilinear couplings of $\zeta_1$ to
the Higgs field and these couplings are proportional to $\lambda_7$, for
$\zeta_1\approx\chi$. By taking $m_{\zeta_1}=$ 200 GeV and $\lambda_7=0.1$, we
have found the following pair-annihilation cross sections for the processes of
Eqs. (\ref{quat}) and (\ref{tri}):
\begin{equation}
\left.\sigma v_{rel}\right|_{t\bar{t}}=0.17~{\rm pb},\quad
\left.\sigma v_{rel}\right|_{WW}=0.94~{\rm pb},\quad
\left.\sigma v_{rel}\right|_{ZZ}=0.44~{\rm pb},\quad
\left.\sigma v_{rel}\right|_{HH}=0.51~{\rm pb}
\label{crss}
\end{equation}
We see that the cross section into $WW$ channel is dominant among all
possible pair-annihilations of $\zeta_1$. The above cross section values
decrease with increasing $m_{\zeta_1}$. From these cross section values, we
see that,
for few hundred GeV of mass to $\zeta_1$ and for $\lambda_7\sim0.1$, using the
processes of Eqs. (\ref{quat}) and (\ref{tri}), the relic abundance
of dark matter can be fitted in this model. However, in this model, the trilinear
coupling of $\zeta_1$ to the Higgs field also drives the process of dark matter scattering
against a nucleus. This process has be searched in many dark matter experiments
\cite{XENON:2018voc,PandaX-4T:2021bab,LZ:2022ufs}, which use xenon as target
nucleus. Since no sign of dark matter is found in these experiments, upper
bounds on the cross section of dark matter against a nucleon have
been obtained. In the Dirac scotogenic model, using \cite{Ma:2021roh},
we estimate the spin-independent cross section of $\zeta_1\approx\chi$ with
xenon nucleus as
\begin{eqnarray}
\sigma_0&=&\frac{1}{\pi}\left(\frac{m_{\zeta_1}m_{Xe}}{m_{\zeta_1}+m_{Xe}}\right)^2
\left|\frac{54f_p+77f_n}{131}\right|^2,
\nonumber \\
\frac{f_p}{m_p}&=&\left[0.075+\frac{2}{27}(1-0.075)\right]
\frac{\lambda_7}{m_{\zeta_1}m_H^2},
\nonumber \\
\frac{f_n}{m_n}&=&\left[0.078+\frac{2}{27}(1-0.078)\right]
\frac{\lambda_7}{m_{\zeta_1}m_H^2}
\end{eqnarray}
Here, $m_p$, $m_n$ and $m_{Xe}$ are masses of proton, neutron and xenon nucleus,
respectively. As stated before, upper bounds on $\sigma_0$ have been set
due to null results of dark matter detection in various experiments. Among
these, the most stringent limit is $\sigma_0<6.5\times 10^{-48}$ cm$^2$
\cite{LZ:2022ufs}. In order to satisfy this limit, in the Dirac scotogenic model,
we get $\lambda_7<4.8\times 10^{-5}$ for $m_{\zeta_1}\sim$ 100 GeV. For this
suppressed value of $\lambda_7$ and for no fine tuning in the mass of $m_{\zeta_1}$,
we see that the cross sections in the
processes of Eqs. (\ref{quat}) and (\ref{tri}) are highly suppressed and we
cannot fit the relic abundance of dark matter in this model. On the other hand,
for a fine tuned mass of $m_{\zeta_1}\approx m_H/2$, via the process
$\zeta_1\zeta_1\to H^*\to b\bar{b}$, the relic abundance of dark matter can be
fitted for $\lambda_7=1.3\times 10^{-5}$, which also satisfies the upper limit
on $\sigma_0$. However, from the indirect searches
of dark matter, thermal annihilation cross section of dark matter into $b\bar{b}$
channel has been ruled out for dark matter mass of up to 300 GeV
\cite{Abazajian:2020tww}. As a result of this, we see that the processes in Eqs.
(\ref{quat}) and (\ref{tri}) cannot consistently explain the dark matter
phenomenology, for the case of $\zeta_1\approx\chi$.

Now, we consider the case $\zeta_1\approx\eta_R^0$, where the dark matter is
dominantly made up of a component of $SU(2)_L$ doublet. In this case, in order to
explain the relic density of dark matter, pair-annihilation of $\zeta_1$
through the processes given in Eqs. (\ref{t-chan})$-$(\ref{tri}) can be analyzed.
It should be noted that the process of Eq. (\ref{t-chan}) happen for the
case $\zeta_1\approx\eta_R^0$ through the $f_{\alpha k}$ couplings of
Eq. (\ref{lag}). As a result of this, pair-annihilation cross section for
this process is analogous to that of Eq. (\ref{paircs}), where $h_{k\alpha}V^*_{\alpha i}$
should be replaced by $f^*_{\alpha k}U^*_{\alpha i}$. Now, we see that, due to small masses
to neutrinos, this pair-annihilation cross section is highly suppressed as
compared to the required amount in order to explain the relic density of
dark matter. The processes in Eqs. (\ref{quat}) and (\ref{tri}) are driven
by the quartic and trilinear couplings of $\zeta_1$ to the Higgs field. We
see that, for the case $\zeta_1\approx\eta_R^0$, these couplings are
proportional to $\lambda_3+\lambda_4+\lambda_5$. As a result of this,
the cross section values of Eq. (\ref{crss}) are also applicable to the case
$\zeta_1\approx\eta_R^0$, where we take $m_{\zeta_1}=$ 200 GeV and
$\lambda_3+\lambda_4+\lambda_5=0.1$. Hence, the relic density of dark matter
can be fitted for the case $\zeta_1\approx\eta_R^0$.
However, analogous to what we
described above, due to non-observation of dark matter in direct detection
experiments, the quantity $\lambda_3+\lambda_4+\lambda_5$ should be
suppressed to around $10^{-5}$. As a result of this, the processes in
Eqs. (\ref{quat}) and (\ref{tri}) cannot consistently explain the relic density of
dark matter, for the case $\zeta_1\approx\eta_R^0$. Apart from the processes
of Eqs. (\ref{t-chan})$-$(\ref{tri}), due to gauge interactions, the
following annihilations are also possible for the case of $\zeta_1\approx\eta_R^0$.
\begin{equation}
\zeta_1\zeta_1\to W^+W^-,\quad \zeta_1\zeta_1\to ZZ,\quad
\zeta_1\zeta_1\to W^+W^-\gamma\gamma
\label{anner}
\end{equation}
In the above equation, the first two annihilations happen due to quartic
interactions and the third annihilation happens due to mediation of $\eta^+$
field at tree level. Moreover, for $m_{\zeta_1}<m_{W,Z}$, one of the $V=W^\pm,Z$ in the
first two processes of Eq. (\ref{anner}) can be off-shell, and thereby, we get
3- and 4-body final states from the annihilation products of
$VV^*$ and $V^*V^*$, respectively. A priori,
for $m_{\zeta_1}\gapprox m_W$, the first process of Eq. (\ref{anner}) can give
the required amount of thermal annihilation cross section in order to fit the
relic density of dark matter.
Moreover, constraints due to indirect detection of dark matter from pair-annihilation
into gauge bosons are weaker \cite{Abazajian:2020tww}. Nevertheless, the
first two processes of Eq. (\ref{anner}) can induce scattering between
$\zeta_1$ and a nucleus at 1-loop level. In fact, there exist other 1-loop
processes, which are mediated by gauge interactions, for the above mentioned
scattering. Although this scattering happens at 1-loop level,
given the strong constraints on this due to direct detection experiments,
we may expect it is challenge for $\zeta_1\approx\eta_R^0$ to evade these
constraints.

\section{Comparision with the scotogenic model}
\label{sadd}

In this section, we compare the phenomenology of Dirac scotogenic model with
that of scotogenic model \cite{Ma:2006km}, where neutrinos are Majorana particles.
It is described in Sec. \ref{s2} that there is an analogy between these two models
in terms of field content and also from the view point of neutrino masses and mixing. Here,
we describe some more analogy between these two models in terms of phenomenological
observables quantities of dark matter, LFV and collider signals.

In the previous section, we have discussed the status of scalar dark matter in
the Dirac scotogenic model. We have discussed two possibilities, where in one
case $\zeta_1\approx\chi$ and in the other case $\zeta_1\approx\eta_R^0$. Now,
below we compare the scalar dark matter phenomenology of this model with that of
scotogenic model \cite{Ma:2006km}. It to remind here that in the scotogenic
model, the singlet field $\chi$ does not exist. Hence, a possibility for scalar dark
matter in the scotogenic model is $\zeta_1=\eta_R^0$. As a result of this,
the phenomenological discussion we have given in the previous section for the
case of $\zeta_1\approx\eta_R^0$ is applicable to the scotogenic model. A difference
in the scotogenic model is that, in addition to the process of Eq. (\ref{t-chan}),
the following process $\zeta_1\zeta_1\to\nu^M_i\nu^M_j$ can also happen, which
violates lepton number. Here, $\nu^M_i$ is a neutrino field of scotogenic model,
which is a Majorana particle. The above process is mediated in the scotogenic
model by the analogous couplings of $f_{\alpha k}$, which are given in Eq. (\ref{lag}).
As already described below Eq. (\ref{lag}), in the scotogenic model, there exist
$N^M_k$ field, instead of $(N_k,N^c_k)$. The field $N^M_k$ is Majorana in the scotogenic
model. As a result of this, the annihilation cross section for the above process
is found to be
\begin{equation}
\sigma(\zeta_1\zeta_1\to\nu^M_i\nu^M_j)v_{rel}=\frac{S}{16\pi}\sum_{k}
\left|\sum_{\alpha,\beta}f_{\alpha k}U_{\alpha i}f_{\beta k}U_{\beta j}\right|^2
\frac{M_K^2}{(M_k^2+m_{\zeta_1}^2)^2}
\end{equation}
Here, $M_k$ is the mass of $N^M_k$ field of the scotogenic model and $S$ is a
symmetry factor which is 1(2) for $i\neq j(i=j)$. For $f_{\alpha k}\sim1$,
$m_{\zeta_1}\sim$ 100 GeV and $M_k=$ 2.78 TeV, the above pair-annihilation
cross section is around 1 pb, which is the required amount in order to fit
the relic abundance of dark matter. Moreover, it looks there exist no constraints
on the above pair-annihilation cross section from experiments. On the other hand,
see \cite{Super-Kamiokande:2020sgt,ANTARES:2019svn,IceCube:2017rdn} for indirect
searches of dark matter, where an upper limit on the
pair-annihilation cross section of dark matter into $\nu\bar{\nu}$ mode is set to
around $10^{-24}$ cm$^3$/s. As a result of the above description,
in the scotogenic model, it is possible to fit the relic density of dark matter and
avoid the indirect detection bounds on it. However, as already described
in the previous section, the first two processes of Eq. (\ref{anner}) can
induce scattering of $\zeta_1$ with a nucleus at 1-loop level. The loop diagrams
for this scattering are driven by gauge couplings and are mediated by SM
fields. Hence, it appears that the amplitude of these diagrams have only the loop
suppression factor. It may be worth to compute the cross section for the above
scattering in order to see if it satisfies the direct detection bound
\cite{LZ:2022ufs} on the dark matter.

As described in Sec. \ref{s2}, there is a region of parameter space where the
couplings $f_{\alpha k}$ can be of order one and these couplings drive
LFV processes \cite{Workman:2022ynf} in the Dirac scotogenic model.
So far none of the LFV processes are
observed in experiments and upper bounds have been set on the branching ratios
of various LFV decays and also on the conversion rate of $\mu$ to $e$ in a
nucleus\cite{Workman:2022ynf}. Among the LFV decays, stringent
limits exist on the branching ratios of $\mu\to e\gamma$ \cite{MEG:2016leq}
and $\mu\to3e$ \cite{SINDRUM:1987nra}.
The above mentioned LFV processes are driven in the Dirac scotogenic model
due to mediation of $\eta^+$ and $N^D_k$ at 1-loop level. Analyzing LFV processes
is out of the scope of this paper. Nevertheless, in our work, experimental
limits on these processes can be satisfied by taking the masses of $\eta^+$ or $N^D_k$
to be sufficiently high, for $f_{\alpha k}\sim1$. In addition to this, there
is also a possibility of suppressing the couplings $f_{\alpha k}$ in order to
satisfy the above experimental limits.

Below, we compare the Dirac scotogenic and scotogenic models in
terms of LFV processes.
As stated before, Feynman diagrams for LFV processes in the Dirac scotogenic
model are driven by $\eta^+$ and $N^D_k$. Whereas, in the scotogenic model \cite{Ma:2006km},
the corresponding Feynman diagrams are driven by $\eta^+$ and $N^M_k$, where
$N^M_k$ is an additional Majorana field.
In the scotogenic model, LFV processes
have been analyzed in \cite{Toma:2013zsa}. We see that, in the Feynman diagrams
for LFV processes of scotogenic model, one should replace the $N^M_k$-propagator with
$N^D_k$-propagator in order to get the corresponding Feynman diagrams of Dirac
scotogenic model. As a result of this, by interchanging the mass of $N^M_k$ with $N^D_k$,
we get the same branching ratio expressions for $\mu\to e\gamma$
in both these models. However, the branching ratio expression for $\mu\to3e$ should
be different in these models, which is explained below. The amplitude for
$\mu\to3e$ arises from the following contributions in both these models:
$\gamma$-penguin, $Z$-penguin, Higgs-penguin and box diagrams. The penguin diagrams
should give same kind of expressions for amplitude in both these models, due to
above mentioned replacement of propagators. On the other hand, the box diagrams
of scotogenic model involve diagrams which are due to Dirac and Majorana
nature of $N^M_k$-propagator. Whereas, these diagrams are driven only due to
Dirac nature of $N^D_k$-propagator in the Dirac scotogenic model. As a result
of this, we get additional contribution to box diagrams in the scotogenic model
as compared to that of Dirac scotogenic model. Now, we
see that, the expression for conversion rate of $\mu$ to $e$ should be same in
both these models, since this conversion happens due to penguin diagrams.

The additional fields in the Dirac scotogenic model are $N^D_k$, $\eta^+$ and
$\zeta_i$, $i=1,2,3$. The best way to test this model in collider experiments is
by probing the $\eta^+$ field. $\eta^+$ can be produced in the LHC experiment
via gauge interactions and it decays as $\eta^+\to\ell^+N^D_1$ or
$\eta^+\to W^+\zeta_1$. Here, $N^D_1$ is the lightest among $N^D_k$. The
above two decays give us missing energy plus a charged lepton or a di-jet signal,
depending on the decay channel of $W$. Analyzing collider signals of this model
is out of the scope of this work. However,
see \cite{vonBuddenbrock:2016rmr,Crivellin:2021ubm},
for some collider analysis on scalar sector in related models. The above described
signal in the Dirac scotogenic model is also possible in the scotogenic model.
The difference between these two models in terms of field content is the presence of the
singlet scalar field $\chi$ in the Dirac scotogenic model. Since the singlet field
does not experience gauge interactions, it is a challenge to probe the existence of
$\chi$ field in the Dirac scotogenic model. Hence, one needs to develop some
techniques to
distinguish the above two models in collider experiments. It is worth to do a
detailed analysis on this aspect in future.

\section{Conclusions}
\label{s8}

In this work, we have studied on the vacuum structure of the scalar potential of
Dirac scotogenic model. One of the motivations of this model is to explain neutrino
masses through a radiative mechanism, and also to have Dirac nature to
the neutrinos. The other motivation is to have a stable dark matter candidate.
After analyzing the scalar potential of this model, we have found that eleven
different minima are possible. Out of these eleven, the $N1$ minimum
of Eq. (\ref{n1}) is the desired minimum of this model. Only with this minimum, the
motivations of this model can be achieved consistently. As a result of this,
we have worked on to see if the $N1$ minimum can be made as the global minimum
of this model. Through our numerical analysis, we have shown that plenty of
parameter space exist where $N1$ is the global minimum. In our numerical
analysis, we have found that the $N1$ minimum can coexist with certain other minima
in some regions of parameter space. However, in the viable parameter space
of this model, we have not found the coexistence of the $N1$ minimum with
charge-breaking minima. We have justified this statement with an analytical
demonstration to it in the appendix \ref{app}.

We have studied the Higgs to diphoton decay in this model, since the scalar
sector of this model has an implication on this decay. After doing a generic
scan over parameters of the scalar potential, we have found that the signal
strength of this decay can be within the experimentally allowed region, but
most likely to be around the lower 3$\sigma$ allowed value of this
quantity. With some tuning of the parameters, the signal strength of this
decay is found to be as high as 1.3. In the numerical analysis, we have found
that the experimentally allowed values of this quantity can be explained
irrespective of the fact that the $N1$ minimum coexist with other minima or not.

Finally, we have studied on the possibility of making the lightest among the
additional scalar particles
of this model, as a candidate for dark matter. We have found that the singlet
scalar field of this model cannot be a viable candidate for dark matter. This
we have found, due to the fact that, the current bounds from the direct and
indirect detection
of dark matter in experiments rule out the possibility of explaining the
relic density of dark matter. The other possibility for scalar dark matter in this
model is the $\eta_R^0$ field,
which is a component of $SU(2)_L$ doublet. In this case, we have found that,
constraints due to direct detection bounds on the dark matter are difficult
to be satisfied.

\appendix
\section{Saddle points}
\label{app}

From the numerical analysis of Sec. \ref{s5}, we have noticed that certain minima
do not coexist with the $N1$ minimum. Below we present analytical calculations
through which we justify why
some of these minima do not coexist with the $N1$ minimum. Our methodology in
these calculations
is based on the discussions given in \cite{Ferreira:2004yd,Ferreira:2019iqb}.

In Sec. \ref{s5}, it is described that, by demanding $\frac{|A|}{v_{EW}}\leq0.1$
in our numerical analysis, we have not
found a region of coexistence between the minima $C11$ and $N1$. Here we show
that, in the limit that $A$ is negligibly small parameter, $C11$ becomes a saddle point
in the region where $N1$ is a minimum. From the minimization conditions of $C11$,
which are given in Tab. \ref{t2}, we notice that $A$ becomes a small variable
if either $\lambda_4+\lambda_5$ or $v_{\phi(11)}v_{\eta(11)}/v_{\chi(11)}$
is suppressed. As a result of this, we neglect terms involving the above mentioned
variables in comparison to other terms of the scalar potential. Now, apart
from satisfying the minimization conditions of $C11$, we need to evaluate the
mass-square eigenvalues of the scalar fields in order to check if $C11$ becomes
a minimum or not. For calculating these eigenvalues, we have described the
parametrization of scalar fields of an SP
in Eq. (\ref{para}). We express these scalar fields in the
following basis: $\varphi=(\phi_R^1,\phi_I^1,\eta_R^1,\eta_I^1,\phi_R^0,
\phi_I^0,\eta_R^0,\eta_I^0,\chi_R)$. Now, the general form of the mixing
mass-square matrix among the scalar fields of an SP is given by
\begin{equation}
[M^2]_{ij}=\frac{\partial^2V}{\partial\varphi_i\partial\varphi_j}=
\frac{\partial^2V}{\partial x_l\partial x_m}\frac{\partial x_l}{\partial\varphi_i}
\frac{\partial x_m}{\partial\varphi_j}+\frac{\partial V}{\partial x_l}
\frac{\partial^2x_l}{\partial\varphi_i\partial\varphi_j}
\label{mas-sqr}
\end{equation}
Here, $i,j=1,\cdots,9$ and $x_l$, where $l=1,\cdots,5$, are defined in Eq. (\ref{bi}).

The last term of Eq. (\ref{mas-sqr}) becomes zero for the case of $C11$, since
after using the minimization conditions, we get $\frac{\partial V}{\partial x_l}=0$.
As a result of this, for the SP $C11$, Eq. (\ref{mas-sqr})
becomes into $M^2_{C11}=Y^TBY$. Here, $Y$ and $B$ are matrices of orders $5\times 9$
and $5\times 5$, respectively. The elements of these matrices are given below.
\begin{equation}
[Y]_{li}=\frac{\partial x_l}{\partial\varphi_i},\quad
[B]_{lm}=\frac{\partial^2V}{\partial x_l\partial x_m}=[M_4]_{lm}+
\frac{\partial^2V_3}{\partial x_l\partial x_m}
\end{equation}
In the limit that $A$ is a small variable, the matrix $B$ takes the following
form.
\begin{equation}
B=\left(\begin{array}{ccccc}
 & & & 0 & 0 \\
 & B_1 & & 0 & 0 \\
 & & & \frac{A}{v_{\chi(11)}} & 0 \\
0 & 0 & \frac{A}{v_{\chi(11)}} & 2(\lambda_4+\lambda_5) & 0 \\
0 & 0 & 0 & 0 & 2(\lambda_4-\lambda_5)
\end{array}\right),\quad
B_1=\left(\begin{array}{ccc}
\lambda_1 & \lambda_3 & \frac{1}{2}\lambda_7 \\
\lambda_3 & \lambda_2 & \frac{1}{2}\lambda_8 \\
\frac{1}{2}\lambda_7 & \frac{1}{2}\lambda_8 & \frac{1}{2}\lambda_6
\end{array}\right)
\end{equation}
Now, as it is argued in \cite{Ferreira:2004yd}, it is possible to express
the matrix $M^2_{C11}$ in the following block form.
\begin{equation}
M^2_{C11}=\left(\begin{array}{cc} 0 & 0 \\ 0 & Y^{\prime^T}BY^\prime \end{array}\right)
\end{equation}
Here, $Y^\prime$ is a $5\times 5$ matrix
which depends only on the VEVs of $C11$. From the above form of $M^2_{C11}$,
we see that $C11$ has four Goldstone bosons. This is expected since the SP $C11$
breaks the electroweak and charge symmetries, which are continuous.
Also, from the above equation, we
see that, if all the eigenvalues of $B$ are positive (negative), then $M^2_{C11}$
is positive (negative) definite, and thus, $C11$ is minimum (maximum). On the
other hand, if $B$ has both positive and negative eigenvalues, then $C11$
becomes a saddle point.
Now, in the limit that $A$ is
a small variable, the first three eigenvalues of $B$ are determined by that of
$B_1$. We see that $Tr(B_1)>0$, since due to the conditions of Eq. (\ref{bfbc}).
As a result of this, $B$ has at least one positive eigenvalue, in the limit
that $A$ is a small variable.

In Eq. (\ref{vc11}), we have given the difference in potential depths at the
stationary points $C11$ and $N1$. In the limit that $A$ is a small variable, the
last term of Eq. (\ref{vc11}) can be neglected. Moreover, in this limiting
process, $m_3^2+\lambda_7v_{EW}^2$ is nearly equal to one of the eigenvalues
of the scalar fields of the SP $N1$. Hence, in a region where $N1$ is a
minimum and in the above limiting process, we get $V_{C11}-V_{N1}>0$. To get
the expression for $V_{C11}-V_{N1}$, we have used the general expression of
Eq. (\ref{difpot}). Here, we see that $X_{N1}^TV^\prime_{C11}=0$. Hence, in
the limit that $A$ is a small variable, we get $V_{C11}-V_{N1}\approx\frac{1}{2}
X_{C11}^TV^\prime_{N1}$. Now, we define $\tilde{X}_{C11}$, $\tilde{X}_{N1}$,
$\tilde{V}^\prime_{N1}$ and $\tilde{M}_2$ as 3-column matrices, whose elements are the
first three elements of $X_{C11}$, $X_{N1}$, $V^\prime_{N1}$ and $M_2$, respectively.
With these definitions and in the above limiting process, we get the following
relation.
\begin{equation}
\left.\frac{\partial V}{\partial x_l}\right|_{C11}=0\implies
\tilde{X}_{C11}\approx-B_1^{-1}\tilde{M}_2
\end{equation}
After using the above relation, we get
\begin{eqnarray}
V_{C11}-V_{N1}\approx\frac{1}{2}X_{C11}^TV^\prime_{N1}&=&\frac{1}{2}
\tilde{X}_{C11}^T\tilde{V}^\prime_{N1}
\nonumber \\
&=&-\frac{1}{2}\tilde{V}^{\prime^T}_{N1}B_1^{-1}\tilde{M}_2
=-\frac{1}{2}\tilde{V}^{\prime^T}_{N1}B_1^{-1}
(\tilde{V}^\prime_{N1}-B_1\tilde{X}_{N1})
\nonumber \\
&=&-\frac{1}{2}\tilde{V}^{\prime^T}_{N1}B_1^{-1}\tilde{V}^\prime_{N1}
\end{eqnarray}
Earlier we have argued that $V_{C11}-V_{N1}>0$ in a region where $N1$ is a minimum.
Hence, after using the above relation, we see that $B_1$ should not be a
positive definite matrix, and thus, one of the eigenvalues of $B_1$ is negative.
As a result of this, in the limit that $A$ is a small variable, one the eigenvalues
of $B$ is negative. Combing this result with the earlier result that
$B$ has at least one positive eigenvalue, we see that $C11$ becomes
a saddle point in a region where $N1$ is a minimum and also that $A$ is negligibly small
variable.

The above described result on the nature of $C11$ is valid even if $A=0$. However,
for $A=0$, we get $\lambda_4+\lambda_5=0$, and thus, the inverse of
$B$ does not exist. To circumvent this problem, we have used $B_1$, which is
a sub-matrix of $B$, in the above described analysis.

The above described analysis can be applied to other SPs of
$N2$, $C9$ and $C10$, in order to show that these become saddle points in the
region where $N1$ is a minimum. The difference we encounter is that the last
term of Eq. (\ref{mas-sqr}) does not vanish for the above mentioned
SPs. As a result of this, we have explicitly computed the mixing masses for the
scalar fields of the above SPs. For instance, in the case of
$C9$, we have found that the fields $\eta_R^1$, $\phi_R^0$ and $\eta_R^0$
mix together, whose masses are given by
\begin{equation}
Z^T\tilde{B}Z,\quad
\tilde{B}=\left(\begin{array}{ccc}
\lambda_2 & \lambda_3 & \lambda_2 \\
\lambda_3 & \lambda_1 & \lambda_3 \\
\lambda_2 & \lambda_3 & \lambda_2
\end{array}\right),\quad
Z=\left(\begin{array}{ccc}
c_{\eta(9)} & 0 & 0 \\
0 & v_{\phi(9)} & 0 \\
0 & 0 & v_{\eta(9)}
\end{array}\right)
\end{equation}
Now, using an analogous formalism described for the case of $C11$, we can show
that the matrix $\tilde{B}$ has one positive and one negative eigenvalue, apart from
a zero eigenvalue. As a result of this, $C9$ becomes a saddle point in a region
where $N1$ is minimum.

In Sec. \ref{s5}, we have mentioned that the minima $N6$ and $N7$ do not
coexist with the $N1$ minimum. Moreover, we have also described that the percentage
of coexistence between the minima $N8$ and $N1$ is zero for $A=0$. All the above mentioned
SPs become saddle points in a region where $N1$ is a minimum.
We have realized this statement through our numerical analysis, but otherwise, we do not
have an analytical proof for this.

\bibliography{refs}
\bibliographystyle{JHEP}
\end{document}